\documentclass[preprint,12pt]{elsarticle}

\usepackage{amssymb}

\usepackage{graphicx}
\usepackage{dcolumn}
\usepackage{bm}
\usepackage{multirow}
\usepackage{float}
\usepackage{romannum}
\usepackage{amsmath}%
\usepackage{MnSymbol}%
\usepackage{wasysym}%
\usepackage{color}%
\usepackage[normalem]{ulem}%
\usepackage{cprotect}

\newcounter{bla}

\expandafter\ifx\csname url\endcsname\relax
  \def\url#1{\texttt{#1}}\fi
\expandafter\ifx\csname urlprefix\endcsname\relax\fi
\expandafter\ifx\csname href\endcsname\relax
  \def\href#1#2{#2} \def\path#1{#1}\fi

\journal{Computer Physics Communications}

\begin{document}

\begin{frontmatter}

\title{MCBTE: A variance-reduced Monte Carlo solution of the linearized Boltzmann transport equation for phonons}

\author[a,b]{Abhishek Pathak}
\author[a]{Avinash Pawnday}
\author[a]{Aditya Prasad Roy}
\author[b]{Amjad J. Aref}
\author[c]{Gary F. Dargush}
\author[a]{Dipanshu Bansal\corref{author}}

 \cortext[author] {Corresponding author.\\\textit{E-mail address:} dipanshu@iitb.ac.in}
 \address[a]{Department of Mechanical Engineering, Indian Institute of Technology Bombay, Powai, Mumbai 400076, India}
 \address[b]{Department of Civil Structural and Environmental Engineering, SUNY Buffalo, Buffalo, NY, 14260, USA}
 \address[c]{Department of Mechanical and Aerospace Engineering, SUNY Buffalo, Buffalo, NY, 14260, USA}

\begin{abstract}
MCBTE solves the linearized Boltzmann transport equation for phonons in three-dimensions using a variance-reduced Monte Carlo solution approach. The algorithm is suited for both transient and steady-state analysis of thermal transport in structured materials with size features in the nanometer to hundreds of microns range. The code is portable and integrated with both first-principles density functional theory calculations and empirical relations for the input of phonon frequency, group velocity, and mean free path required for calculating the thermal properties. The program outputs space- and time-resolved temperature and heat flux for the transient study. For the steady-state simulations, the frequency-resolved contribution of phonons to temperature and heat flux is written to the output files, thus allowing the study of cumulative thermal conductivity as a function of phonon frequency or mean free path. We provide several illustrative examples, including ballistic and quasi-ballistic thermal transport, the thermal conductivity of thin films and periodic nanostructures, to demonstrate the functionality and to benchmark our code against available theoretical/analytical/computational results from the literature. Moreover, we parallelize the code using the Matlab Distributed Computing Server, providing near-linear scaling with the number of processors.

\end{abstract}


\end{frontmatter}



{\bf PROGRAM SUMMARY}

\begin{small}
\noindent
{\em Program Title:} MCBTE                                       \\
{\em CPC Library link to program files:} (to be added by Technical Editor) \\
{\em Developer's repository link:} https://github.com/abhipath90/MCBTE \\
{\em Code Ocean capsule:} (to be added by Technical Editor)\\
{\em Licensing provisions(please choose one):} GPLv3  \\
{\em Programming language:} MATLAB                              \\
{\em Nature of problem:} Calculation of time- and space-dependent temperature and heat flux profiles, and frequency-resolved effective thermal conductivity in structured systems where heat is carried by phonons\\
{\em Solution method:} Solution of linearized Boltzmann transport equation for phonons, variance-reduced Monte Carlo approach\\
{\em Runtime:} About 1 to 10 hours on a personal computer
   \\

\end{small}


\section{Introduction}
The size of electronic components such as the gate size of a transistor is aggressively scaled down. In modern CPUs, the gate size is of the order of $10$\,nm, and efforts are directed towards bringing it down to 1\,nm~\cite{desai2016mos2}. At high processing speeds, the gates generate a significant quantity of heat, which needs to be dissipated quickly to prevent failure from overheating. To facilitate this effort, it is essential to have an efficient and reliable method to simulate the heat conduction of electronic devices that can capture the phenomenon at the nano to micro length scale and the hundreds of femtosecond (fs) to nanosecond (ns) time scale~\cite{chen2005nanoscale, cahill2014nanoscale}. 

In the nonmagnetic semiconducting crystalline solids such as silicon, lattice vibrations are primary heat carriers~\cite{chen2005nanoscale,cahill2014nanoscale,ziman2001electrons}, where a quanta of lattice vibration is referred to as a phonon. In contrast to metals, electron contribution in semiconductors is small/negligible at room temperature. In such solids, under external perturbation, for instance, heating of one end of a one-dimensional (1D) object, the drift of phonons leads to deviation from the thermodynamic equilibrium, which is restored by phonon-phonon (ph-ph) scattering~\cite{mazumder2001monte}. The number of such ph-ph scattering events would be large enough at the macro-scale to restore thermodynamic equilibrium and the transport is diffusion-like. This diffusion-like transport is adequately described using the Fourier law of heat conduction~\cite{chen2005nanoscale, cahill2014nanoscale}. Phonons would also scatter from impurities and sample boundaries, but these scattering events do not necessarily restore thermal equilibrium, as the scattered phonon has the same energy and polarization (transverse/longitudinal, acoustic/optic) as the incident phonon~\cite{chen2005nanoscale, ziman2001electrons, mazumder2001monte}. On the other hand, if the characteristic dimension of the sample is smaller than the mean distance traveled by phonons [i.e., mean free path (MFP) of phonons], ph-ph scattering events would be few, and the thermal equilibrium would not be restored. In such a scenario, the Fourier law is not adequate to model the heat conduction in the sample~\cite{chen2005nanoscale, cahill2014nanoscale}, and alternative models must be sought.

\begin{figure}[]
  \centering
  \includegraphics[width=0.5\textwidth]{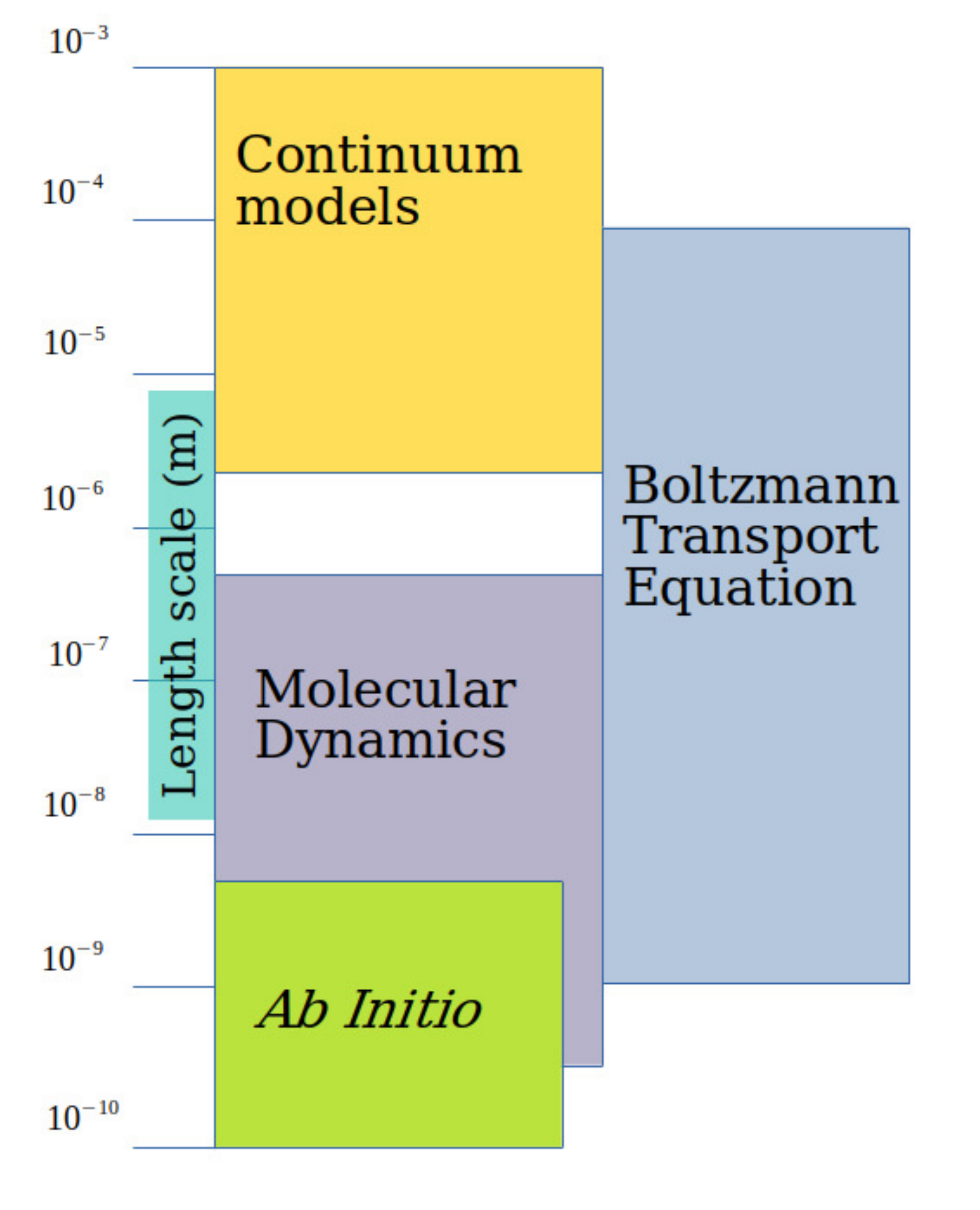}  
  \caption{Range of applicability of various heat conduction modeling techniques.}
  \label{fig:range_methods}
\end{figure}

The Boltzmann transport equation (BTE) is oft-used in the modeling of the heat conduction where the Fourier law breaks down, and can adequately describe the equilibrium and non-equilibrium phenomenon~\cite{chen2005nanoscale, mazumder2001monte, peraud2011efficient, peraud2012alternative, peraud2014monte}. BTE treats phonons as particles, and wave properties of phonons are not considered. Figure~\ref{fig:range_methods} shows the domain of applicability in terms of length scale for various methods used for modeling heat conduction problems. BTE is applicable from continuum to nanoscale.  Thus, it is ideal for the simulation of systems and devices that are too small for continuum models to be useful and too large for an all-atom description using molecular dynamics or first-principles based methods. The vast application domain of BTE has led to considerable efforts towards analytical and numerical solutions of BTE for crystalline solids under simplifying assumptions and geometries. The solution techniques fall into two categories based on their approach: (1) deterministic methods~\cite{narumanchi2004submicron,majumdar1993microscale,chai1994finite,ravishankar2010finite}; (2) stochastic or Monte Carlo (MC) methods. One of the advantages of deterministic methods is that they converge fast and provide good control over the statistical uncertainty of the results. A detailed description of deterministic methods is presented in Ref.~\citenum{mittal2011prediction}. However, for complex geometries and highly anisotropic heat transport, deterministic methods require high fidelity spatial and angular discretization. MC methods are desirable for such cases and alleviate the computational challenge associated with the high dimensionality of the distribution function and stability problems in simulating the advection process~\cite{baker2005variance}. 

One of the first MC schemes for solving BTE was developed by Klitsner \emph{et al.}~\cite{klitsner1988phonon} to study low-temperature heat conduction. In this scheme, internal scattering was neglected, which allowed them to simulate the ballistic limit only. Peterson included the ph-ph scattering in his MC scheme under the relaxation-time approximation (RTA) along with a simplified assumption of the Debye solid~\cite{peterson1994direct}. Mazumder and Majumdar~\cite{mazumder2001monte} built upon Peterson's work and presented the comprehensive solution approach with minimal simplifying assumptions. For example, transverse and longitudinal phonons were explicitly included in their scheme. Lacroix \emph{et al.}~\cite{lacroix2005monte} included frequency-dependent MFP and developed a distribution function that satisfied energy conservation during phonon scattering events. Hao \emph{et al.}~\cite{hao2009frequency} introduced a method to apply periodic boundary conditions in the MC simulation. In recent years, in a series of papers~\cite{peraud2011efficient, peraud2012alternative, peraud2014monte}, Hadjiconstantinou and co-workers developed a variance reduced formulation for recasting BTE in the deviational energy form and linearized it for cases when only a small deviation from the equilibrium temperature is expected in the domain. Our code is based on their MC scheme derived for linearized deviational energy-based BTE (LBTE)~\cite{peraud2012alternative, peraud2014monte}.  The code takes phonon properties, for example, phonon energy, group velocity, and lifetime, as input either from empirical models or from a first-principles based deterministic solution of phonon BTE in crystalline materials such as calculated using PhonTS~\cite{chernatynskiy2015phonon}, Phono3py~\cite{phono3py}, AlmaBTE~\cite{carrete2017almabte} and Alamode~\cite{tadano2014anharmonic}. However, in contrast to the first-principles based deterministic solution approaches, our code uses these phonon properties to simulate steady-state and transient thermal transport in complex 3D nanoscale geometries. Moreover, using our code, various phonon scattering lifetimes (impurity, boundary, Umklapp, normal) can be treated independently, instead of a single relaxation time using Matthiessen's rule, to study their combined effect on thermal transport, as we demonstrate later in the manuscript.

In what follows, we first briefly describe the theory of LBTE and the MC solution before describing the details of the implementation in our code. The rest of the paper is organized as follows: Section~\ref{sec_theory} describes the basic theory of BTE for phonons and derivation of LBTE. In Section~\ref{sec_algo_lin}, we describe various steps involved in the MC simulation of LBTE in detail. Section~\ref{sec_code} describes the input and output files for our code. We benchmark the output of our code against analytical expressions and literature data in Section~\ref{sec_bench} and demonstrate parallelization in Section~\ref{sub_parallel}. We summarize the potential applications of the code in Section~\ref{sec_summary}.

\section{Theory} \label{sec_theory}
Below we briefly describe the BTE for phonons and derivation of LBTE. A more detailed description can be found in Ref.~\citenum{peraud2014monte}. 

\subsection{Boltzmann transport equation (BTE)} \label{sub_BTE_origin}
BTE is a conservation equation for classical particles in the position and velocity phase space, $(\mathbf{x},\mathbf{v})$, originally formulated for the kinetic description of dilute gases~\cite{chen2005nanoscale}. BTE describes the evolution of single-particle probability distribution function $f(\mathbf{x},\mathbf{v}, t)$ at time $t$,
\begin{equation}
  \label{eq:BTE_gas_kin}
  \frac{\partial f}{\partial t} + \mathbf{v} \cdot \nabla_{\mathbf{x}} f + \mathbf{F} \cdot \nabla_{\mathbf{v}} f = \left .\frac{\partial f}{\partial t} \right \rvert_{coll}.
\end{equation}
Here, $\mathbf{F}$ is the external force acting per unit mass of gas molecules. The physical interpretation of BTE is that the change in  $f(t,\mathbf{x},\mathbf{v})$ due to the advection term of the left-hand side and collision term on the right-hand side is always balanced in a volume element in phase space. For the hard-sphere model of gases, the collision operator is of the following form~\cite{cercignani1988boltzmann},
\begin{equation}
  \label{eq:coll_gas}
  \left .\frac{\partial f}{\partial t} \right \rvert_{coll} = \int \int [f'_1f' - f_1f] \; ||\mathbf{v - v_1}|| \; \sigma d^2\Omega d^3\mathbf{v_1}.
\end{equation}
Here $\sigma=\hat{d}^2/4$ is the differential cross-section for hard spheres, $\hat{d}$ is the effective diameter of gas molecules, and $f'=f(t,\mathbf{x,v'}), f_1=f(t,\mathbf{x,v_1}), f'_1=f(t,\mathbf{x,v'_1})$; $[{\mathbf{v_1,v}}]$ are pre-collision velocities and $[{\mathbf{v'_1,v'}}]$ are the post-collision velocities, related through the scattering angle $\Omega$. Integration in velocities is over all possible velocities in phase space and integration over solid angles is over the entire surface of the unit sphere.

\subsection{BTE for phonons} \label{sub_Phonon_BTE}
The quantum mechanical description of a phonon has both wave and  particle nature. If we neglect the phonon coherence effects, then in the absence of external force, BTE is adapted to produce a semi-classical description of the evolution of the phase space density distribution for dilute `phonon-gas',
\begin{equation}
  \label{eq:phonon_BTE}
  \frac{\partial f}{\partial t} + \mathbf{\nabla_k} \omega(\mathbf{k},p)\cdot \mathbf{\nabla_x}f = \left . \frac{\partial f}{\partial t} \right \rvert_{coll}.
\end{equation}
Since phonons are not affected by an external force, $\mathbf{F} \cdot \nabla_{\mathbf{v}} f$ is dropped from Eq.~\eqref{eq:BTE_gas_kin}. Phonon frequency $\omega$ is related to its wave vector $\mathbf{k}$ through the dispersion relation $\omega(\mathbf{k},p)$, where $p$ denotes the phonon polarization. The equilibrium solution of Eq.~\eqref{eq:phonon_BTE} at temperature $T$ is given by the Bose-Einstein distribution
\begin{equation}
  \label{eq:feq}
  f^{eq}(\omega,T) = \frac{1}{\exp \left ( \frac{\hbar \omega}{k_BT}\right) -1}.
\end{equation}
The significant difference between the hard-sphere model for dilute gases and that of phonons is in the nature of their scattering events. In the hard-sphere model, energy and momentum are always conserved in all scattering events. However, a phonon may or may not conserve momentum during a scattering event. When a phonon is scattered through a  ``two-phonon process'' (for example, by an impurity), its momentum changes  ($\mathbf{k \neq k'}$), but the energy remains unchanged. A ``three-phonon process'' occurs when either two phonons combine to create a third phonon (type \Romannum{1} process) or a phonon decays into two phonons (type \Romannum{2} process). The following conservation equations apply to the three phonon processes,
\begin{equation}
  \label{eq:3_phn_conserve}
  \mathbf{k \pm k' = k'' + H } \quad \quad \mathrm{(type \; \Romannum{1}/\Romannum{2} \; processes)}, \text{ and}
\end{equation}
\begin{equation}
  \label{eq:3_phn_conserve_2}
  \omega \pm \omega' = \omega'' \quad \quad \mathrm{(type \; \Romannum{1}/\Romannum{2} \; processes)}.
\end{equation}
For the normal processes $\mathbf{H}=0$,  while for the Umklapp processes $\mathbf{H=G}$, where $\mathbf{G}$ is the reciprocal lattice vector. Umklapp scattering does not conserve momentum and is the primary source of direct resistance to heat conduction. Higher-order processes such as the ``four-phonon process'' are believed to be negligible at low and moderate temperatures and are usually neglected\cite{ecsedy1977thermal}. Recently, it is argued that four-phonon processes are comparable to three-phonon processes at medium to high-temperature range and  contribute significantly to anharmonic materials~\cite{feng2017four,feng2016quantum}. However, calculating scattering rates for four-phonon processes is still a computational challenge, and we neglect the higher-order processes in what follows. The scattering operator, considering only two and three phonon processes, is written as \cite{pascual2009thermal}
\begin{equation}
  \label{eq:coll_term_full}
  \begin{split}
    \left . \frac{\partial f}{\partial t} \right \rvert_{coll} &=  \sum_{\mathbf{k'},p'} {f_{\mathbf{k'}p'}(f_{\mathbf{k}p} + 1) - f_{\mathbf{k}p}(f_{\mathbf{k'}p'} + 1)} \mathcal{Q}^{\mathbf{k'}p'}_{\mathbf{k}p} \\
    & + \sum_{\mathbf{k'},p'\mathbf{k''},p''} {(f_{\mathbf{k}p} + 1)(f_{\mathbf{k'}p'} + 1) f_{\mathbf{k''}p''} - f_{\mathbf{k}p}f_{\mathbf{k'}p'}(f_{\mathbf{k''}p''} + 1)} \mathcal{Q}^{\mathbf{k''}p''}_{\mathbf{k}p,\mathbf{k'}p'} \\
    & + \frac{1}{2} \sum_{\mathbf{k'},p'\mathbf{k''},p''} {(f_{\mathbf{k}p} + 1)f_{\mathbf{k'}p'} f_{\mathbf{k''}p''} - f_{\mathbf{k}p}(f_{\mathbf{k'}p'} + 1) (f_{\mathbf{k''}p''} + 1)} \mathcal{Q}^{\mathbf{k'}p',\mathbf{k''}p''}_{\mathbf{k}p},
  \end{split}
\end{equation}
where $\mathcal{Q}$ is the transition probability matrix of a phonon with wave-vector ${\bf k}$ and polarization $p$ to  another state denoted by ${\bf k'}p'$ and/or ${\bf k''}p''$, usually a nonlinear function of wave-vector $\mathbf{k}$. Any physical observable at any position is calculated by summing up all the phonon contributions in the wave-vector space. For isotropic systems with dense energy levels, these summations are converted to integrals using the density of states (DOS)
\begin{equation}
  \label{eq:DOS}
  D(\omega,p)=\frac{k^2(\omega,p)}{2 \pi V_g(\omega,p)},
\end{equation}
where $V_g(\omega,p)\equiv ||\mathbf{\nabla_x}\omega(\mathbf{k},p)||$ is the phonon group velocity. We note that the $D(\omega,p)$ is measured using inelastic neutron/x-ray scattering or can also be directly calculated from first-principles phonon simulations without simplifying assumptions. Using $D(\omega,p)$, the number density $n(t,x)$ of phonons is written as
\begin{equation}
  \label{eq:number_density}
  n(t,x) = \sum_p \int \int \int f(t,\mathbf{x},\omega,\theta,\phi,p) \frac{D(\omega,p)}{4 \pi} \sin \theta\,d\omega\,d\theta\,d\phi,
\end{equation}
where $\theta$ and $\phi$ are the polar and azimuthal angles in spherical coordinates. Similarly, the energy density $U(t,\mathbf{x})$ and the heat flux $\mathbf{q}(t,\mathbf{x})$ are given by,
\begin{equation}
  \label{eq:ener_dens}
  U(t,\mathbf{x}) = \sum_p \int \hbar \omega f(t,\mathbf{x},\omega,\theta,\phi,p) \frac{D(\omega,p)}{4 \pi} \sin \theta \,d\omega\,d\theta\,d\phi,\text{ and}
\end{equation}
\begin{equation}
  \label{eq:flux}
  \mathbf{q}(t,\mathbf{x}) = \sum_p \int \hbar \omega \mathbf{V}_g f(t,\mathbf{x},\omega,\theta,\phi,p) \frac{D(\omega,p)}{4 \pi} \sin \theta \,d\omega\,d\theta\,d\phi,
\end{equation}
respectively. For a system far from thermodynamic equilibrium, the temperature $T(t,{\bf x})$, at  position $\mathbf{x}$ is defined by equating $U(t,\mathbf{x})$ with the energy density obtained from the equilibrium distribution as,
\begin{align}
  \label{eq:temp_define}
  \sum_p \int \hbar \omega f(t,\mathbf{x},\omega,\theta,\phi,p) &\frac{D(\omega,p)}{4 \pi} \sin \theta \,d\omega\,d\theta\,d\phi \nonumber \\
  &= \sum_p \int \hbar \omega f^{eq}(\omega,T(t,{\bf x}))D(\omega,p)\,d\omega.
\end{align}

\subsection{Relaxation time approximation} \label{sub_RTA}
The solution of BTE is challenging due to the non-linearity of the collision operator. To solve BTE, the collision operator is usually approximated using simplified models. One of the frequently used approaches is the `relaxation time approximation' (RTA)~\cite{ziman2001electrons, chen2005nanoscale}. RTA assumes that the role of all ph-ph scattering events is to drive (`relax') the system to the local equilibrium $f^{loc}$.  For a constant relaxation time ($\tau$), this leads to
\begin{equation}
  \label{eq:relax_approx}
  \left . \frac{\partial f}{\partial t} \right \rvert_{coll} = -\frac{f - f^{loc}}{\tau}.
\end{equation}
For brevity, we have omitted the explicit dependence of $f$ on various parameters. In the rarefied gas dynamics literature, this model is referred to as the Bhatnagar-Gross-Krook model~\cite{bhatnagar1954model}. To capture the complexities of the ph-ph scattering for different $\omega$, $p$, and $T$, $\tau$ is usually written as $\tau(\omega,p,T)$. Hence, under the RTA approximation, Eq.~\eqref{eq:phonon_BTE} reduces to
\begin{equation}
  \label{eq:BTE_relax}
  \frac{\partial f}{\partial t} + \mathbf{\nabla_k} \omega(\mathbf{k},p)\cdot \mathbf{\nabla_x}f = \frac{ f^{loc} -f}{\tau(\omega, p , T)}.
\end{equation}
Here, $T$ in $\tau(\omega,p,T)$ is calculated using Eq.~\eqref{eq:temp_define}. For phonons, RTA implies that the scattering step consumes the phonons at a rate of $1/\tau(\omega,p,T)$ and generates new phonons from the local equilibrium distribution at the same rate. The newly generated phonons are independent of the consumed phonons. However, the energy conservation demands that the total energy of the newly generated phonons is equal to the total energy of consumed phonons. The term $f^{loc}$ is the Bose-Einstein distribution at pseudo-temperature $T_{loc}$, where, $T_{loc}$ is calculated using the energy equivalence of the consumed and newly generated phonons during the scattering process as 
\begin{align}
  \label{eq:T_local}
  \sum_p \int \frac{\hbar \omega f(\mathbf{x},\omega,\theta,\phi,p,t)}{\tau(\omega,p,T)} & \frac{D(\omega,p)}{4 \pi} \sin \theta \, d\omega \, d\theta \, d\phi \nonumber \\
  &= \sum_p \int \frac{\hbar \omega f^{eq}(\omega,T_{loc})}{\tau(\omega,p,T)} D(\omega,p) \, d\omega.
\end{align}

\subsection{Deviational energy-based BTE} \label{sub_dev_BTE}
The conventional MC solution applied to Eq.~\eqref{eq:BTE_relax} does not strictly  satisfy energy conservation during the scattering process, and the energy fluctuates around its mean value. The fluctuations add to the statistical uncertainty of the measured quantity and can create a bias by interacting with other fluctuating processes of the system~\cite{peraud2014monte}. Recasting BTE in the energy form explicitly  satisfies the  energy conservation~\cite{peraud2011efficient}. If we multiply Eq.~\eqref{eq:BTE_relax} by $\hbar \omega$ and define $e=\hbar \omega f$ and $e^{loc} = \hbar \omega f^{loc}$, we obtain the energy-based BTE
\begin{equation}
  \label{eq:energy_BTE}
  \frac{\partial e}{\partial t} + \mathbf{V_g} \cdot \mathbf{\nabla_x} e = \frac{e^{loc}-e}{\tau(\omega,p,T)}.
\end{equation}
 Here, $\mathbf{V_g} = \mathbf{\nabla_k} \omega(\mathbf{k},p)$ is the phonon group velocity. In this formulation, each computational particle represents a fixed amount of energy $e=\hbar \omega f$ and a strict energy conservation is achieved by conserving the number of particles. Another limitation of the conventional MC simulation is a low signal-to-noise (S/N) ratio when departure from the equilibrium distribution is small~\cite{hadjiconstantinou2003statistical}. This is a typical manifestation in the calculation of  effective thermal conductivity ($\kappa_{eff}$) of periodic nano structures~\cite{jeng2008modeling}. Low S/N ratio can be improved by simulating only the deviation of energy distribution from the equilibrium~\cite{peraud2011efficient} using the control variate technique of the variance reduction. In this technique, the equilibrium energy distribution $e^{eq}_{T_{eq}} = \hbar \omega f^{eq}(\omega,T_{eq})$ is a control variate, and we solve for the deviation from equilibrium $e^d = e - e^{eq}_{T_{eq}}$. Recasting Eq~\eqref{eq:energy_BTE} in the deviational form leads to 
\begin{equation}
  \label{eq:deviational_BTE}
  \frac{\partial e^d}{\partial t} + \mathbf{V_g} \cdot \mathbf{ \nabla_x} e^d = \frac{e^{loc} - e^{eq}_{T_{eq}} - e^d}{\tau(\omega, p , T)}.
\end{equation}
The recasting of Eq.~\eqref{eq:energy_BTE} into Eq.~\eqref{eq:deviational_BTE} assumes that $T_{eq}$ is independent of ${\bf x}$ and $t$. For $(T-T_{eq})\ll T_{eq}$, control variate $e^{eq}_{T_{eq}}$ is close to the actual distribution $e$, thus providing an ideal condition for variance reduction. In addition, the moments of equilibrium distribution are known semi-analytically and their effects are deterministically added to the results to improve the computational efficiency.  

\subsection{Linearization of deviational energy-based BTE for phonons} \label{sec_lin_RTA}
Deviational energy-based BTE (Eq.~\eqref{eq:deviational_BTE}) can be further simplified by linearization for cases where the deviation from equilibrium is small. Under the approximation -- $(T-T_{eq}) \ll T_{eq}$ and $(T_{loc}-T_{eq}) \ll T_{eq}$, using the Taylor series expansion we write
\begin{equation}
  \label{eq:lin_scatt}
  \frac{e^{loc} - e^{eq}_{T_{eq}}}{\tau(\omega,p,T)} = \frac{1}{ \tau(\omega,p,T_{eq})} \frac{d e^{eq}_{T_{eq}}}{dT} (T_{loc}-T_{eq}) + \mathcal(O) \left( \frac{T_{loc} - T_{eq}}{T_{eq}} \right)^2.
\end{equation}
Using Eq.~\eqref{eq:lin_scatt}, Eq.~\eqref{eq:deviational_BTE} is written as
\begin{equation}
  \label{eq:lin_BTE}
  \frac{\partial e^d}{\partial t} + \mathbf{V_g} \cdot \nabla_{\mathbf{x}} e^d = \frac{\mathfrak{L}(e^d)- e^d}{\tau(\omega,p,T_{eq})},
\end{equation}
where
\begin{equation}
  \label{eq:lin_operator}
  \mathfrak{L}(e^d) = (T_{loc} - T_{eq}) \frac{d e^{eq}_{T{eq}} }{dT}.
\end{equation}
$T_{loc}$ is defined using the same energy equivalence as Eq.~\eqref{eq:T_local}. Under deviational formulation $T_{loc}$ modifies to
\begin{equation}
  \label{eq:pseudo_deviational}
  \begin{aligned}
    \int_{\omega}\sum_{p}\frac{D(\omega,p)(e^{loc} - e^{eq}_{T_{eq}})}{\tau(\omega,p,T)} d \omega &= \int \int \int  \sum_p \frac{1}{4\pi}\frac{D(\omega,p)e^d}{\tau(\omega,p,T)} \sin \theta\,d\omega\,d\phi\,d\theta.
  \end{aligned}
\end{equation} 
Equation~\eqref{eq:pseudo_deviational} is simplified under linearization as follows
\begin{align}
  \label{eq:lin_energy}
  (T_{loc} - T_{eq})\int_{\omega} \sum_p &\frac{D(\omega, p)}{\tau(\omega,p,T_{eq})} \frac{de^{eq}_{T_{eq}}}{dT}\,d \omega \nonumber \\
  &= \int \int \int \sum_p \frac{1}{4\pi}\frac{D(\omega,p) e^d}{\tau(\omega,p,T_{eq})} \sin \theta\,d \theta\,d\phi\,d\omega,
\end{align}
where $\frac{de^{eq}_{T_{eq}}}{dT}$ is calculated using the analytical expression given by
\begin{equation}
  \label{eq:BE_derivative}
  \frac{de^{eq}_{T_{eq}}}{dT} = k_B \left ( \frac{\hbar \omega }{2 k_B T_{eq}} \right )^2 \frac{1}{\sinh^2 \left ( \frac{\hbar \omega }{k_B T_{eq}}\right )}.
\end{equation}
Post scattering, new deviational particles  are drawn from the distribution $ \mathfrak{L}(e^d)/\tau(\omega,p,T_{eq})$, which after normalization and using Eq.~\eqref{eq:lin_energy} becomes
\begin{equation}
  \label{eq:norm_post_coll}
  \frac{\frac{D(\omega, p) \Delta t}{4 \pi} \frac{\frac{de^{eq}_{T_{eq}}}{dT}}{\tau(\omega,p, T_{eq})}}{\int\limits_{\omega}\sum\limits_p \frac{D(\omega, p) \Delta t}{4 \pi} \frac{\frac{de^{eq}_{T_{eq}}}{dT}}{\tau(\omega,p, T_{eq})} \, d\omega}.
\end{equation}
Normalized post-scattering distribution in Eq.~\eqref{eq:norm_post_coll} is independent of $T$ and $T_{loc}$, hence, the scattering process does not require their calculation. The Monte Carlo (MC) implementation of LBTE is simplified due to this independence from $T$ and $T_{loc}$ as we discuss later. 

\subsection{Spatially varying control} \label{sub_spatial}
The above discussion of deviational BTE was limited to constant $T_{eq}$ in ${\bf x}$ and $t$. In the control variate formulation, it is well-known that a control closer to the non-equilibrium distribution will increase the effectiveness of variance reduction . Hence, a location dependent $T_{eq}$ is preferable  in defining appropriate control. If we take $T_{eq}$ as an explicit function of ${\bf x}$ (but independent of $t$), Eq.~\eqref{eq:deviational_BTE} is written as 
\begin{equation}
  \label{eq:spatial_BTE}
  \frac{\partial e^d}{\partial t} +  \mathbf{V_g} \cdot \nabla_{\mathbf{x}} e^d = \frac{\left [ e^{loc} - e^{eq}_{T_{eq}(\mathbf{x})}\right ] - e^d}{\tau(\omega , p , T)} - \mathbf{V_g} \cdot \nabla_{\mathbf{x}}T_{eq}({\bf x}) \frac{d e^{eq}_{T_{eq}(\mathbf{x})}}{d T}.
\end{equation}
The implication of spatially varying control function is the appearance of an additional source term on the right-hand side. Now, if we linearize Eq.~\eqref{eq:spatial_BTE} using a constant $T_{eq,0}$ within the range of $T_{eq}({\bf x})$, we obtain
\begin{equation}
  \label{eq:lin_spatial}
  \frac{\partial e^d}{\partial t} + \mathbf{V_g} \cdot \nabla_{\mathbf{x}} e^d = \frac{\mathfrak{L}(e^d)- e^d}{\tau(\omega,p,T_{eq,0})} - \mathbf{V_g} \cdot \nabla_{\mathbf{x}}T_{eq}({\bf x}) \frac{d e^{eq}_{T_{eq,0}(\mathbf{x})}}{d T},
\end{equation}
Linearization with respect to constant $T_{eq,0}$ (in contrast to $T_{eq}({\bf x})$) preserves all advantages of spatially constant control presented in Section~\ref{sec_lin_RTA}, while keeping the same order of approximation. In the following section we describe the MC solution of LBTE given by Eqs.~\eqref{eq:lin_BTE} and~\eqref{eq:lin_spatial}.

\section{Monte Carlo solution of LBTE} \label{sec_algo_lin}
The solution of linearized deviational energy-based BTE (LBTE) for phonons using an MC approach generates samples from the initial deviational energy distribution $e^d(\mathbf{x},t=0)$. The samples are propagated (i.e., drift and scatter) via the governing equation(s) to collect the statistics in $(\mathbf{x},t)$  for estimating the new distribution $e^d(\mathbf{x},t)$. This $e^d$ is subsequently used to calculate the physical observables, such as temperature and heat flux. Here, $e^d$ is sampled by $N$ computational particles using 
\begin{align}
  \label{eq:deviational_approx}
  e^d(t,\mathbf{x},\omega,\theta,\phi,p) &\frac{D(\omega,p)}{4 \pi} \nonumber \\
  &= \mathcal{E}^{d}_{eff}\sum_i s_i \delta^3(\mathbf{x}- \mathbf{x_i})\delta(\omega - \omega_i)\delta(\theta - \theta_i) \delta(\phi - \phi_i)\delta_{p,p_i}, 
\end{align}
where $s_i$ is the sign of a computational particle given by the sign of $e^d = e - e^{eq}_{T_{eq}}$. Since $e^d$ can be positive or negative, $s_i$ is an essential parameter.  A particle having negative $e^d$ will decrease the temperature and flux (i.e., the flux will be in the opposite direction to its travel direction). This behavior is tracked by the parameter $s_i$. $\mathcal{E}^{d}_{eff}$ is the fixed amount of effective deviational energy carried by a computational particle and is calculated at the start of the simulation. The dynamics of the particles is governed by the LBTE and makes use of the direct simulation Monte Carlo (DSMC) method developed by Bird~\cite{bird1963approach}. In DSMC, BTE is solved by discretization in time, where each time integration step is split into a collision-less advection sub-step and a subsequent scattering sub-step~\cite{wagner1992convergence,bird1994molecular}. A detailed discussion of DSMC based MC solution of Eq.~\eqref{eq:BTE_relax} is presented in Ref.~\citenum{peraud2014monte}. In what follows, we describe the numerical implementation and solution of LBTE.

\subsection{Effective deviational energy} \label{eff_dev}
The total deviational energy $E^d_{tot}$ is calculated by combining the contributions of all the sources. The sources include -- initial conditions, volumetric heat source, and isothermal boundaries. The deviational energy associated with the $i^{th}$ source is given by
\begin{equation}
  \label{eq:dev_tot}
  E^d_i = \sum_{p} \int_t \int_{\omega} \int_{\phi}  \int_{\theta} \int_V \frac{D}{4\pi}  |Q_i| dV\,\sin (\theta) \, d \theta \, d \phi \, d \omega \, dt .
\end{equation}
Where, $Q_i$ is the phase-space energy density associated with the $i^{th}$ source, and $dV$ is the differential volume element in the position space. If the number of sources is $N_s$, $E^d_{tot}$ is written as $E^d_{tot}= \sum_i^{N_s}E^d_i$. Note that the magnitude of $E^d_i$ is calculated here. The expressions of $Q$ for different sources included in our implementation are~\cite{peraud2014monte},
\begin{equation}
  \label{eq:source_energy}
  \begin{aligned}
    &Q_{init} = \delta (t)(T_{init} - T_{eq}) \frac{d e^{eq}_{T{eq}} }{dT} \quad(\text{Initial conditions}) \\
    &Q_{bnd} = \delta(\mathbf{x}) H( \mathbf{V_g} \cdot \hat{\bf{n}}) ( \mathbf{V_g} \cdot \hat{\bf{n}}) (T_{b}-T_{eq}) \frac{d e^{eq}_{T{eq}} }{dT} \quad (\text{Isothermal boundary}) \\
    &Q_{volumesource} = - \mathbf{V_g} \cdot \nabla_{\mathbf{x}}T_{eq}({\bf x}) \frac{d e^{eq}_{T_{eq}(\mathbf{x})}}{d T} \quad (\text{Spatially varying control})
  \end{aligned}
\end{equation}
Here, $T_{init}$ is the initial temperature, $H$ is the Heaviside function specifying the direction of particle emission inside the domain, $\hat{\bf{n}}$ is the inward normal to the isothermal boundary, and $T_{b}$ is the boundary temperature.  The effective deviational energy is calculated as $\mathcal{E}^d_{eff}=E^d_{tot}/N$, where $N$ is the number of computational particles in the simulation. Furthermore, the normalized cumulative deviational energy for the ordered list of sources is calculated as $E^d_{cum}(i) = \sum_1^i E^d_i$. 

\subsection{Initialization} \label{sub_init}
The probability for a particle to originate from the $i^{th}$ source is equal to $E^d_i/E^d_{total}$. To choose the source of origin of the particle, a random number $\mathfrak{R_{1}} \in [0,1)$ is drawn. If $E^d_{cum}(i) \leqslant \mathfrak{R_{1}} < E^d_{cum}(i+1)$, the particle is emitted from the $i^{th}$ source. If the particle is emitted from the initial conditions specified at $t = 0$, the starting time is $t_0 = 0$. Otherwise, the starting time for a particle is chosen by drawing a random number $\mathfrak{R_{2}} \in [0,1)$, and assigning $t_0 = \mathfrak{R_{2}} t_{max}$. Here, $t_{max}$ is the total simulation time. Although steady-state simulations do not have time as a variable in the formulation, a pseudo-time is used to model the  dynamics of the particles and all particles are assigned $t_0=0$.

\subsubsection{Emission from initial conditions} \label{emit_init}
The initial position of the particle is assigned based on the sampling of volume $V$ of the simulation domain.  The simulation domain volume is discretized into volumetric elements (hereafter referred to as a spatial cell). The probability of a particle originating from one of the spatial cells is proportional to the total deviational energy of the spatial cell, which is chosen following the same procedure as described above for selecting a particular source. The size and number of such cells in the simulation domain are dependent upon the desired spatial resolution. Further sampling of position in a cell depends on its shape. For a rectangular orthogonal hexahedron, a simple uniform sampling of all three components of the position vector $\mathbf{x}$ is sufficient. An arbitrary 3D domain can be represented using a tetrahedron as a building block. A detailed procedure of such uniform sampling is presented in Ref.~\citenum{mazumder2001monte}. An analytical sampling of the spectral domain is usually more challenging. Numerically, it is done as follows. Material data such as ${\bf V_g}$ and $\tau$ are usually sampled at an equidistant discrete point in the spectral domain. We treat those sample points as $\omega_{0,i}$, and distance between them as $\delta \omega$. The number of phonons in the $i^{th}$ bin is calculated as
\begin{equation}
  \label{eq:Nbin}
  \begin{split}
    \frac{ N(\omega_{0,i})}{V} &= \frac{1}{\mathcal{E}^d_{eff}}\sum_p \int | e^d_{init}(\omega_{0,i},T_{init})| \frac{D(\omega,p)}{4 \pi} \sin \theta\,d\omega\,d\theta\,d\phi\\
    &= \frac{1}{\mathcal{E}^d_{eff}}\sum_p |e^d_{init}(\omega_{0,i},T_{init})| D(\omega_{0,i},p) \delta \omega.
  \end{split}
\end{equation}
Here $e^d_{init}$ is the deviational distribution for initial temperature given by
\begin{align}
  \label{eq:dev_init}
      e^d_{init}= \hbar \omega &\left( \frac{1}{\exp(\hbar \omega / k_B T_{init})-1} - \frac{1}{\exp(\hbar \omega / k_B T_{eq})-1}\right) \nonumber \\
      &\approx (T_{init} - T_{eq}) \frac{d e^{eq}_{T{eq}} }{dT}
\end{align}
A uniform random number $\mathfrak{R_3} \in [0,1)$ is drawn to choose a bin. If $F_{j-1} \leq \mathfrak{R_3} < F_j$, the particle is assigned to $j^{th}$ bin, where the cumulative distribution $F_i$ is given by
\begin{equation}
  \label{eq:cumul_freq}
  F_i =  \frac{\sum^i_{j=1}N(\omega_{0,j})}{\sum_{j}N(\omega_{0,j})}.
\end{equation}
The particle is assigned the frequency of the chosen bin. In general, $D(\omega_{0,i},p)$ obtained from experiments and simulations is summed over $p$, and we do not have to explicitly choose the polarization at a given $\omega_{0,i}$. However, if $D(\omega_{0,i},p)$ is given as a function of $p$, we can draw a random number to choose the polarization following the procedure described in Ref.~\citenum{mazumder2001monte}. The sign of the particle is assigned the same as that of $(T_{init}-T_{eq}) $. The traveling direction of the particle is sampled such that each point at the surface of a unit sphere has the same probability. Consequently, the following probability distributions for the polar and azimuthal angles are chosen.
\begin{equation}
  \label{eq:prob_angle}
  P_{\theta} = \frac{\sin \theta}{2} \quad\forall \theta \in [0,\pi) \quad \mathrm{and} \quad P_{\phi} = \frac{1}{2 \pi} \quad\forall \phi \in [0,2 \pi),
\end{equation}
where $\theta$ and $\phi$ are generated from
\begin{equation}
  \label{eq:angle_gen}
  \theta = \cos^{-1}(1-2 \mathfrak{R_4}) \quad \mathrm{and} \quad \phi = 2 \pi \mathfrak{R_5}.
\end{equation}
Here $\mathfrak{R_4}$ and $\mathfrak{R_5}$ are also uniformly distributed random numbers in [0,1).

\subsubsection{Emission from isothermal boundary} \label{emit_bnd}
The position of the particle is assigned based on a uniform sampling of the boundary surface. For a rectangular surface, a uniform sampling along the two orthogonal sides is sufficient. If the boundary is a different polygon, it can be uniformly sampled by representing it with triangle elements, as described in Ref.~\citenum{mazumder2001monte}. Isothermal boundaries emit particles into the simulation domain from an equilibrium distribution of their temperature $T_b$ given as
\begin{equation}
  \label{eq:dev_bnd}
      e^d_b= \hbar \omega \left( \frac{1}{\exp(\hbar \omega / k_B T_b)-1} - \frac{1}{\exp(\hbar \omega / k_B T_{eq})-1}\right) \approx  (T_{b}-T_{eq}) \frac{d e^{eq}_{T{eq}} }{dT}. 
\end{equation}
The number of phonons emitted from the isothermal boundary in the $i^{th}$ frequency bin is given by
\begin{equation}
  \label{eq:iso_flux}
  \begin{split}
    N_b(\omega_{0,i}) & =  \frac{\mathcal{A}}{\mathcal{E}^d_{eff}}\sum_p \int \mathbf{V_g}(\omega_{0,i},p)\cdot {\bf \hat{n}} \, |e^d_b(\omega_{0,i})| \frac{D(\omega_{0,i},p)}{4 \pi} \sin \theta\,d\theta\,d\phi\,\delta\omega\\
    & = \frac{\mathcal{A}}{4\mathcal{E}^d_{eff}} \sum_p  V_g(\omega_{0,i},p) | e^d_b(\omega_{0,i})| D(\omega_{0,i},p)\, \delta \omega,
  \end{split}
\end{equation}
where $V_g = ||\mathbf{V_g}||$ is the magnitude of the phonon group velocity, ${\bf \hat{n}}$ is the unit normal to the boundary pointing inward, $\mathcal{A}$ is the area of the isothermal boundary. The cumulative distribution of Eq~\eqref{eq:cumul_freq} is now calculated using $N_b(\omega_{0,i})$. The frequency of the particle is assigned following the same procedure as followed for the particle emission from initial conditions. The sign of the particle is assigned same as that of $(T_b-T_{eq})$. The traveling direction is sampled from new probability distributions as now directions are uniformly distributed on a hemisphere ($\bm{\mathrm{k\cdot \hat{n}}} > 0 $) instead of the whole unit sphere,
\begin{equation}
  \label{eq:wall_dir}
  P_{\theta} = 2 \cos \theta \sin \theta \quad\forall \theta \in [0, \pi/2) \quad P_{\phi} = 1/(2 \pi) \quad\forall \phi \in [0,2\pi),
\end{equation}
where $\theta$ and $\phi$ are generated using  -- $\theta = \cos^{-1}(\sqrt{\mathfrak{R_6}})$ and $\phi = 2 \pi \mathfrak{R_7}$. Here, $\mathfrak{R_6}$ and $\mathfrak{R_7}$ are uniform random numbers in [0,1).

\subsubsection{Emission from volumetric source} \label{emit_src}
If the particle is emitted from a constant intensity volumetric source, all positions are equally likely within the domain. Consequently, the position is assigned in the same manner as for the particle emission from initial conditions. For spatially varying control (Section~\ref{sub_spatial}), the thermal gradient leads to the particle emission within the body from the following distribution
\begin{equation}
  \label{eq:dev_ss}
  e^d_{ss} =  - \mathbf{V_g} \cdot \nabla_{\mathbf{x}}T_{eq}({\bf x}) \frac{d e^{eq}_{T_{eq}(\mathbf{x})}}{d T}.
\end{equation}
The number of particles emitted in the $i^{th}$ frequency bin is given by
\begin{equation}
  \label{eq:grad_flux}
  \begin{aligned}
   N_{ss}(\omega_{0,i}) &= \frac{1}{\mathcal{E}^d_{eff}}\sum_p \int  |e^d_{ss}(\omega_{0,i},T_i)| \frac{D(\omega,p)}{4 \pi} \sin \theta\,d\omega\,d\theta\,d\phi\,dV\\
    & = \frac{1}{\mathcal{E}^d_{eff}}\sum_p \int | \mathbf{V_g} \cdot \nabla_{\mathbf{x}}T_{eq}({\bf x}) | \frac{d e^{eq}_{T_{eq}(\mathbf{x})}}{d T}  \frac{D(\omega,p)}{4 \pi} \sin \theta\,d\omega\,d\theta\,d\phi\,dV.
  \end{aligned}
\end{equation}
For the case of a uniform thermal gradient, Eq~\eqref{eq:grad_flux} simplifies to
\begin{equation}
  \label{eq:grad_flux_sim}
   N_{ss}(\omega_{0,i}) = \frac{V}{4 \mathcal{E}^d_{eff}} \sum_p   V_g(\omega_{0,i},p) |\nabla_{\mathbf{x}}T_{eq}({\bf x})| D(\omega_{0,i},p)\,\delta \omega.
\end{equation}
The frequency of the particle is assigned following the same procedure as described in Section~\ref{emit_init} and~\ref{emit_bnd}. The traveling direction is now distributed uniformly on a unit hemisphere ($\mathbf{k} \cdot \nabla_{\bf x}T_{eq}({\bf x})>0$ ). The traveling direction is sampled following the same procedure as followed in Section~\ref{emit_bnd}. For $s_i$, since both signs ($+ \text{ or } -$) are equally likely, we draw a random number $\mathfrak{R_8}$. If $\mathfrak{R_8}<0.5$, we assign a positive sign to the particle; otherwise, a negative sign is assigned. Since the particle with a negative sign carries negative flux, we reverse the traveling direction to be consistent with the direction of the flux.

\subsection{Advection and time to next scattering event} \label{sub_advect}
In the advection sub-step, LBTE is solved without the scattering term, i.e., 
\begin{equation}
  \label{eq:advection_eq}
  \frac{\partial e^d}{\partial t} + \mathbf{V_g} \cdot \nabla_{\mathbf{x}} e^d = 0.
\end{equation}
A computational particle travels ballistically, and its position is updated using $\mathbf{x_1} = \mathbf{x_0} + \mathbf{V_g} \Delta t$. Here $\Delta t$ is the time to the next scattering event for which the particle travels uninterrupted unless it encounters a boundary. To calculate $\Delta t$, we solve for scattering sub-step given by the following equation
\begin{equation}
  \label{eq:scatt_eq}
  \frac{\partial e^d}{\partial t} = \frac{\mathfrak{L}(e^d)- e^d}{\tau(\omega,p,T_{eq})}.
\end{equation}
Assuming $\mathfrak{L}(e^d)$ to be constant between $t$ and $t + \Delta t$, we can integrate Eq.~\eqref{eq:scatt_eq} to obtain
\begin{equation}
  \label{eq:scat_step}
  e^d(t + \Delta t) = e^d(t) + (\mathfrak{L}(e^d) - e^d(t)) \left (  1 - \exp \left ( \frac{- \Delta t}{\tau(\omega,p,T_{eq})}\right )\right ).
\end{equation}
Numerical solution of Eq.~\eqref{eq:scat_step} requires that we replace the current particle (which was drawn from $e^d$) with a new particle drawn from $\mathfrak{L}(e^d)$ with a probability
\begin{equation}
  \label{eq:scat_prob}
  P(\omega,p,T_{eq}) =  1- \exp \left ( \frac{-\Delta t}{\tau(\omega,p,T_{eq})}\right )
\end{equation}
By inverting Eq.~\eqref{eq:scat_prob} and replacing $1-P(\omega,p,T_{eq})$ with an uniform random number $\mathfrak{R_9} \in [0,1)$, we get
\begin{equation}
  \label{eq:time_to_scat}
  \Delta t = -\tau(\omega,p,T_{eq}) \ln (\mathfrak{R_9}).
\end{equation}
As opposed to the frequently used Matthiessen rule that combines various scattering processes (i.e., impurity scattering, normal processes, and Umklapp processes) by summing the inverse of their relaxation times $\tau_i$ following $\tau^{-1} = \sum_i\tau_i^{-1} $; in our simulation, we treat three-phonon processes separately from two-phonon processes. We draw two separate time-to-scattering using Eq.~\eqref{eq:time_to_scat}: $\Delta t_2$, for two phonon processes and $\Delta t_3$, for three phonon processes, using their respective values of $\tau$. The particle will undergo scattering at time $\Delta t = \min(\Delta t_2, \Delta t_3)$ and the scattering time is updated as $t_1 = t_0 + \Delta t$. Now three cases arise as particle drifts. 
\begin{enumerate}
\item[(i)] If $\Delta t = \Delta t_3$, the particle is redrawn from the distribution described in Section~\ref{sub_three_phonon}. In this case both $\Delta t_2$ and $\Delta t_3$ are resampled for the newly drawn particle.
\item[(ii)] If $\Delta t = \Delta t_2$, the particle remains the same but its traveling direction is redrawn as described in Section~\ref{sub_two_phonon}. The time $t_1-t_0$ is subtracted from $\Delta t_3$, a new value of $\Delta t_2$ is sampled using Eq.~\eqref{eq:time_to_scat}, and the particle continues to drift and scatter for the remaining time.
\item[(iii)] Between $t_0$ and $t_1$, the particle may encounter a boundary.  For boundary scattering, the segment $(\mathbf{x_0},\mathbf{x_1}]$ is checked for interactions with simulation domain boundaries. If the particle interacts with a boundary at ${\bf x}_b$,  $\mathbf{x_1}$ is set to $\mathbf{x_1} = {\bf x}_b$, and the time of the scattering event is updated as $t_1 = t_0 + ||{\bf x}_b-{\bf x}_0||/||\mathbf{V_g}||$. We note that the computational particle remains the same after the boundary scattering. We subtract $t_1-t_0$ from $\Delta t_2$ and $\Delta t_3$ to calculate the remaining time before the next scattering event.
\end{enumerate}

\subsection{Sampling} \label{sub_sample}
The solution of LBTE does not require the computation of any thermodynamic observable such as temperature for simulation to proceed. Hence the sampling for calculating an observable is performed in the spatial domain where the data is required. In a transient simulation, measurement times are also specified. If the particle is found at any predetermined spatial locations at the measurement times, its contribution to the thermodynamic observables is calculated. If $\Xi(t) = \sum_p \int D/(4 \pi) \, \xi e^d(t) \sin (\theta) \, d \theta \, d \phi \, d \omega \, dV$ is the macroscopic property (i.e., thermodynamic observable) in terms of a general microscopic property $\xi = \xi(\mathbf{x},\omega,p,\theta, \phi)$, then using Eq.~\eqref{eq:deviational_approx}, the contribution of the particle to the macroscopic quantity is calculated as
\begin{equation}
  \label{eq:sampling}
  \tilde{\Xi}(t) = \mathcal{E}^d_{eff}\sum_{i=1}^N s_i \xi(\mathbf{x}_i(t),\omega_i(t),p_i(t),\theta_i(t),\phi_i(t)).
\end{equation}
Equation~\eqref{eq:sampling} is then added to the equilibrium baseline value to get the true estimate of the thermodynamic observable. In our implementation, we calculate temperature and heat flux as follows. If the particle is present at the time of measurement in a sampling volume $V$, its contribution to the energy density is $s_i \mathcal{E}^d_{eff}/V$. The temperature difference from the equilibrium baseline is calculated by dividing the energy density with the heat capacity $C$, i.e., $T_{dev}=s_i \mathcal{E}^d_{eff}/CV$. Similarly, the $x$-component (or $y$ or $z$) of the heat flux is calculated as $q_x=s_i \mathcal{E}^d_{eff} V_{g,x}/V$, where $V_{g,x}$ is the $x$-component of the particle velocity.

\subsection{Scattering and boundary conditions} \label{sub_scatt}
The ballistic drift of the particle is interrupted by phonon scattering (i.e., ph-ph, boundary, impurity). Different scattering processes affect the post-scattering trajectory of the particle differently. In the following, we describe the implementation of various scattering events.

\subsubsection{Two phonon processes} \label{sub_two_phonon}
This type of scattering happens when a particle scatters from an impurity. The impurity randomizes the direction of travel of the particle. Other particle properties remain the same. To simulate the two phonon processes, we draw a new traveling direction of the particle using the procedure described in Section~\ref{emit_init}.

\subsubsection{Three phonon processes} \label{sub_three_phonon}
In three phonon processes, either two phonons combine to create one phonon or one phonon disintegrates into two phonons. For the particle, frequency is drawn from the post-scattering distribution $\frac{\mathfrak{L}(e^d)}{\tau(\omega,p,T_{eq})}$. The number of deviational particles in the $i^{th}$ frequency bin for a given arbitrary time duration $t_{scat}$ is given by 
\begin{equation}
  \label{eq:n_scatt}
  \begin{split}
    \frac{ N_{scat}(\omega_{0,i})}{V} &= \frac{t_{scat}}{\mathcal{E}^d_{eff}}\sum_p \int \frac{\mathfrak{L}(e^d)}{\tau(\omega,p,T_{eq})} \frac{D(\omega,p)}{4 \pi} \sin \theta\,d\omega\,d\theta\,d\phi\\
    &= \frac{ t_{scat} (T_{loc}-T_{eq})}{\mathcal{E}^d_{eff}}\sum_p \frac{\frac{de^{eq}_{T_{eq}}}{dT}}{\tau(\omega,p, T_{eq})} D(\omega_{0,i},p) \delta \omega.
  \end{split}
\end{equation}
Although Eq.~\eqref{eq:n_scatt} depends on $T_{loc}$ and $t_{scat}$, the cumulative distribution (Eq.~\eqref{eq:cumul_freq}) is independent of both. The frequency is assigned following the same procedure as described in Section~\ref{emit_init}. The traveling direction of the particle is also re-sampled following the same procedure as in Section.~\ref{emit_init}. The sign of the particle remains unchanged.

\subsubsection{Adiabatic boundary} \label{sub_adiab}
Adiabatic boundaries reflect the incident particle into the simulation domain. An adiabatic boundary is of two types, namely -- specular and diffusive. In the specular reflection, the outgoing wave-vector $\mathbf{k'}$ is related to the incoming wave-vector $\mathbf{k}$ by
\begin{equation}
  \label{eq:specular}
  \mathbf{k'} = \mathbf{k} - 2(\mathbf{k\cdot\hat{n}})\mathbf{\hat{n}}
\end{equation}
Particle energy and polarization remain unchanged. In the diffusive reflection, the direction of the reflected particle is randomized. We re-sample the traveling direction using Eq.~\eqref{eq:wall_dir}. In practice, the real boundary properties may lie between the specular and diffusive reflection, and is dependent on phonon wavelength as recently experimentally demonstrated on freestanding silicon membranes~\cite{ravichandran2018spectrally}. For such cases, the degree of specularity ($d \in [0,1]$) is defined as the probability of a boundary to behave as a specular mirror. We choose specular and diffusive reflection by drawing a uniform random number $\mathfrak{R_{10}} \in [0,1)$. If $\mathfrak{R_{10}} < d$, the particle is reflected specularly, otherwise diffusively.  Since, experimentally determined $d$ for a particular phonon wavelength is an (unknown) probability distribution~\cite{ravichandran2018spectrally}, drawing a uniform random number may not always be appropriate.

\subsubsection{Isothermal boundary} \label{sub_iso}
Deviational particle incident on an isothermal boundary thermalizes with the boundary, and its deviational energy becomes zero. An isothermal boundary acts as an absorbing boundary for a deviational particle.

\subsubsection{Periodic boundary} \label{sub_period}

\begin{figure}[]
  \centering
  \includegraphics[width=0.7\textwidth]{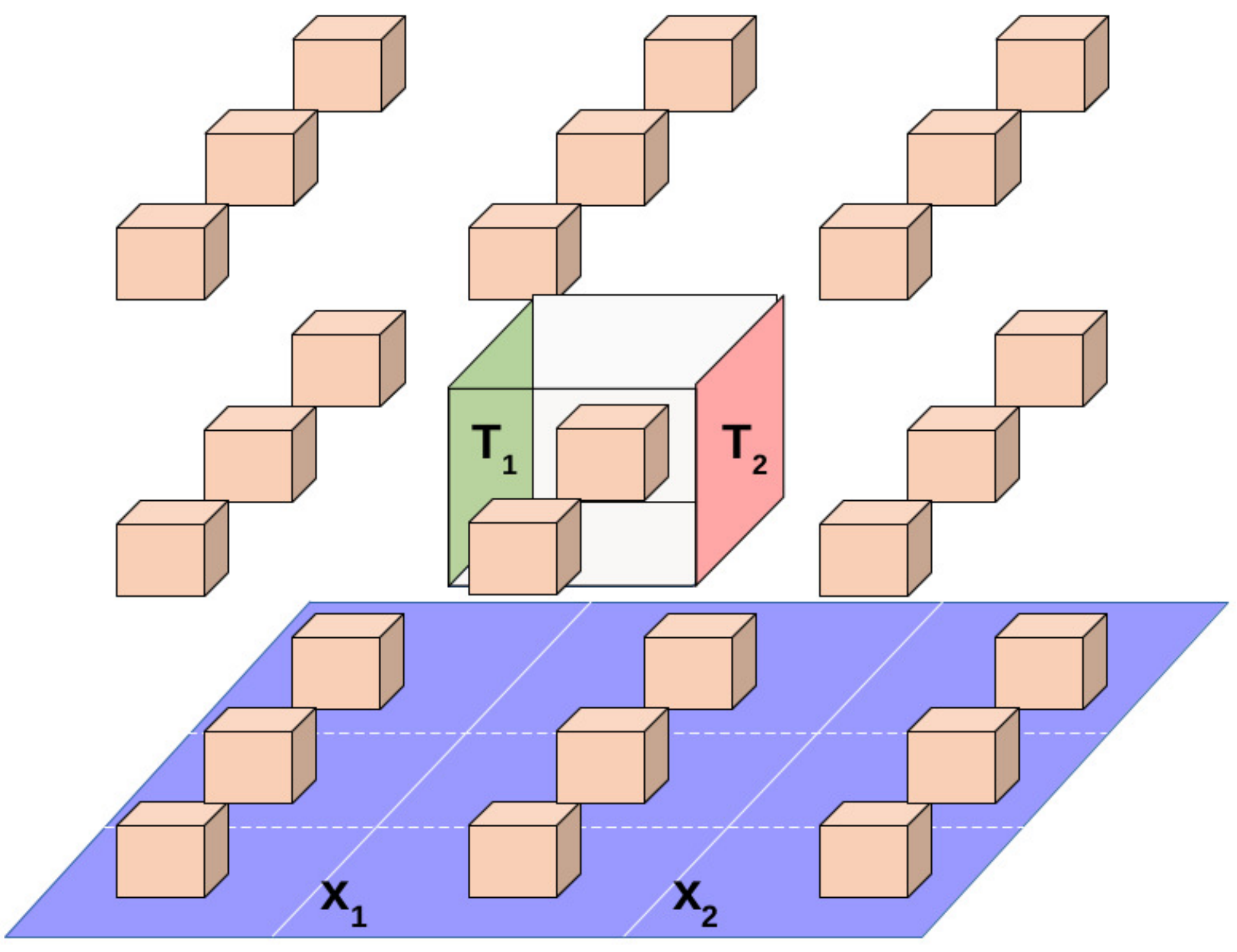}
  \caption{The periodic arrangement of nano inclusions in a matrix. Black lines define the unit cell. For a given constant flux, $T_1$ and $T_2$ are temperature on opposite faces of unit cell at ${\bf x_1}$ and ${\bf x_2}$, respectively. Image adapted from Peraud and Hadjiconstantinou~\cite{peraud2014monte}.}
  \label{fig:periodic_unit}
\end{figure}

For the simulation of periodic nanostructures, we need to introduce periodic boundary conditions in the formulation. A typical implementation of periodic boundary conditions would impose periodicity in the temperature. However, in general, temperature periodicity is not physical. Hence, a constant thermal gradient is applied along the direction of periodicity. In such a scenario, the deviation of phonon distribution from the local equilibrium is periodic~\cite{jeng2008modeling,hao2009frequency}. As shown in Figure~\ref{fig:periodic_unit}, if the periodic boundary pair located at $\bm{\mathrm{x_1}}$ and $\bm{\mathrm{x_2}}$ has local temperatures $T_1$ and $T_2$, respectively, the periodicity of local equilibrium is written as
\begin{equation}
  \label{eq:periodic_ener}
  \begin{aligned}
    &  e^{in}_2 - e^{eq}_{T_2} = e^{out}_1 - e^{eq}_{T_1}, \text{ and} \\
    &  e^{in}_1 - e^{eq}_{T_1} = e^{out}_2 - e^{eq}_{T_2}.
  \end{aligned}
\end{equation}
Here, the first equation describes the case when a particle leaves the domain at $\bm{x_1}$, and the second equation, when it exits the domain at $\bm{x_2}$.  For a fixed control, i.e.\ $T_{eq}$ is independent of $\bf{x}$ and $t$, Eq.~\eqref{eq:periodic_ener} is written in the deviational form as
\begin{equation}
  \label{eq:periodic_dev}
  \begin{aligned}
    &  e^{in}_2 - e^{eq}_{T_{eq}} = e^{out}_1 -e^{eq}_{T_{eq}} + e^{eq}_{T_2}- e^{eq}_{T_1}, \text{ and} \\
    &  e^{in}_1 - e^{eq}_{T_{eq}} = e^{out}_2 - e^{eq}_{T_{eq}} + e^{eq}_{T_1} - e^{eq}_{T_2}
  \end{aligned}
\end{equation}
When a particle is incident on a periodic boundary, it is reinserted from the other side. New particles are generated from the distribution
\begin{equation}
  \label{eq:peri_dev_dis}
  (e^{eq}_{T_1} - e^{eq}_{T_2}) \frac{D(\omega,p)}{4\pi}V_g(\omega,p).
\end{equation}
The spatially variable control simplifies the treatment of periodic boundary conditions when a small thermal gradient is applied in the direction of periodicity. Starting from Eq.~\eqref{eq:periodic_dev}, and linearizing it near $T_{lin}$ (the temperature at which we want to calculate the heat flux and thermal conductivity), we obtain
\begin{equation}
  \label{eq:lin_periodic}
  \begin{aligned}
    & e^{in}_2 - e^{eq}_{T_{lin}} = e^{out}_1 - e^{eq}_{T_{lin}} + \frac{de^{eq}_{T_{lin}}}{dT} (T_2 - T_1) \\
    & e^{in}_1 - e^{eq}_{T_{lin}} = e^{out}_2 - e^{eq}_{T_{lin}} + \frac{de^{eq}_{T_{lin}}}{dT} (T_1 - T_2). \\
  \end{aligned}
\end{equation}
Using $T_{eq}(\mathbf{x}) = T_0 + \mathbf{x}\cdot \nabla_{\mathbf{x}}T_{eq}$, where $T_0$ is a constant, Eq.~\eqref{eq:lin_periodic} becomes
\begin{equation}
  \label{eq:lin_periodic_grad}
  \begin{aligned}
    & e^{in}_2 - e^{eq}_{T_{lin}} = e^{out}_1 - e^{eq}_{T_{lin}} + \frac{de^{eq}_{T_{lin}}}{dT} \nabla_{\mathbf{x}}T_{eq} \cdot (\mathbf{x}_2 - \mathbf{x}_1) \\
    & e^{in}_1 - e^{eq}_{T_{lin}} = e^{out}_2 - e^{eq}_{T_{lin}} + \frac{de^{eq}_{T_{lin}}}{dT} \nabla_{\mathbf{x}}T_{eq} \cdot (\mathbf{x}_1 - \mathbf{x}_2) \\
  \end{aligned}
\end{equation}
Note that the control function under linearization approximation is  $e_{control}=e^{eq}_{T_{lin}} + \mathbf{x} \cdot \nabla_{\mathbf{x}}T_{eq}de^{eq}_{T_{lin}}/dT$. Using $e_{control}$ and rearranging the terms in Eq.~\eqref{eq:lin_periodic_grad} we get
\begin{equation}
  \label{eq:lin_periodic_grad_2}
  \begin{aligned}
    & e^{in}_2 - \left ( e^{eq}_{T_{lin}} + \frac{de^{eq}_{T_{lin}}}{dT} \nabla_{\mathbf{x}}T_{eq} \cdot \mathbf{x}_2 \right ) = e^{out}_1 - \left ( e^{eq}_{T_{lin}} + \frac{de^{eq}_{T_{lin}}}{dT} \nabla_{\mathbf{x}}T_{eq} \cdot \mathbf{x}_1 \right )  \\
    & e^{in}_1 - \left ( e^{eq}_{T_{lin}} + \frac{de^{eq}_{T_{lin}}}{dT} \nabla_{\mathbf{x}}T_{eq} \cdot \mathbf{x}_1 \right ) = e^{out}_2 - \left ( e^{eq}_{T_{lin}} + \frac{de^{eq}_{T_{lin}}}{dT} \nabla_{\mathbf{x}}T_{eq} \cdot  \mathbf{x}_2 \right ),
  \end{aligned}
\end{equation}
which can be written as
\begin{equation}
  \label{eq:periodic_lin_final}
  \begin{aligned}
    & e^{in,d}_2 = e^{out,d}_1 \\
    & e^{in,d}_1 = e^{out,d}_2.
  \end{aligned}
\end{equation}
Equation~\eqref{eq:periodic_lin_final} implies that by using spatially variable control, deviational particles leaving one boundary are inserted from the other boundary without changing their properties.

\subsubsection{Termination} \label{sub_term}
If $t_1 > t_{max}$ or if an isothermal boundary absorbs the particle, its trajectory is terminated. For a steady-state simulation of the periodic domain, there may be no isothermal boundaries to absorb the particles. Any particle incident on a periodic boundary is re-inserted in the domain. In this case, the particle's trajectory is terminated after it has undergone a predefined number of relaxation events (i.e., three phonon processes). The number of relaxation events depends on the problem type, and a convergence study is usually performed to find its appropriate value.

\subsection{Steady-state sampling} \label{sub_steady}
Steady-state is achieved by running simulations with some initial conditions for long enough time. Sampling is done by further running the simulation past that time. The linearized solution scheme presented here makes it possible to directly sample the steady-state solution without explicitly collecting data for the entire duration. If time to reach the steady-state is $t_{ss}$, the steady-state estimate of a macroscopic quantity $\Xi$ can be obtained by time-averaging of Eq.~\eqref{eq:sampling},
\begin{align}
  \label{eq:steady_average}
  \tilde{\Xi}(ss) &= \frac{1}{\mathcal{T}} \int_{t'=t_{ss}}^{t_{ss}+\mathcal{T}} \tilde{\Xi}(t') dt' \nonumber\\
  &= \frac{\mathcal{E}^d_{eff}}{\mathcal{T}} \sum_{i=1}^N s_i \int_{t'=max(t_i^{start}, t_{ss})}^{t_{ss}+\mathcal{T}}\xi(\mathbf{x}_i(t),\omega_i(t),p_i(t),\theta_i(t),\phi_i(t))\,dt'.
\end{align}
Here, $\mathcal{T}$ is the time for which average is calculated beyond $t_{ss}$, and $t_i^{start}$ is the emission time of particle $i$. Since all sources are time-independent in the steady-state (or their influence dies off with time), integrating Eq.~\eqref{eq:dev_tot} in time and adding contribution from all the sources leads to
\begin{equation}
  \label{eq:dev_steady_rate}
  \dot{E}^d_{tot}=\frac{E^d_{tot}}{t_{ss}+\mathcal{T}} = \int_{\omega} \int_{\phi} \int_{\theta} \int_V \frac{D}{4\pi}\left |\sum_j Q_j \right | \sin (\theta) \, d\theta \, d\phi \, d\omega \, dV.
\end{equation}
If we extend the time integration in Eq.~\eqref{eq:steady_average} to $\mathcal{T} \to \infty$, i.e., the time when particle exits the simulation, we obtain
\begin{equation}
  \label{eq:steady_estimate}
  \tilde{\Xi}(ss) = \sum_{i=1}^{N} \dot{\mathcal{E}}^d_{eff} \int_{t'=t_i^{start}}^{t_i^{end}} \xi(\mathbf{x}_i(t),\omega_i(t),p_i(t),\theta_i(t),\phi_i(t))\,dt'.
\end{equation}
Here, $\dot{\mathcal{E}}^d_{eff} = \dot{E}^d_{tot}/N$ is the effective deviational energy rate. If the macroscopic quantity is temperature in a volume $V$, Eq.~\eqref{eq:steady_estimate} simplifies to
\begin{equation}
  \label{eq:temp_steady}
  T_{dev} = \frac{\dot{\mathcal{E}}^d_{eff} }{CV} \sum_i s_i \frac{l_i}{V_{g,i}}.
\end{equation}
Here, $l_i$ is total absolute length traveled by particle $i$. A true estimate of the temperature is calculated by adding the equilibrium value, i.e., $T = T_{eq} + T_{dev}$. Similarly, if the macroscopic quantity is the $x$-component of the heat flux, we get
\begin{equation}
  \label{eq:flux_x_steady}
  q_x = \frac{\dot{\mathcal{E}}^d_{eff} }{V} \sum_i s_i l_{x,i}
\end{equation}
Here, $l_{x,i}$ is the displacement of particle $i$ along the $x$ direction. Since $q$ is zero when $T_{eq}$ is independent of ${\bf x}$, $q_x$ is the true estimate.

\section{MATLAB code I/O} \label{sec_code}
In this section, we present the I/O of MATLAB code of our implementation of the MC solution of LBTE. The linearized algorithm is embarrassingly parallel, and the code uses MATLAB distributed computing server (MDCS) to utilize as many compute nodes as are assigned. The input files of the code are as follows.

\subsection{Material data} 
\begin{figure}
  \centering
  \includegraphics[width=\textwidth]{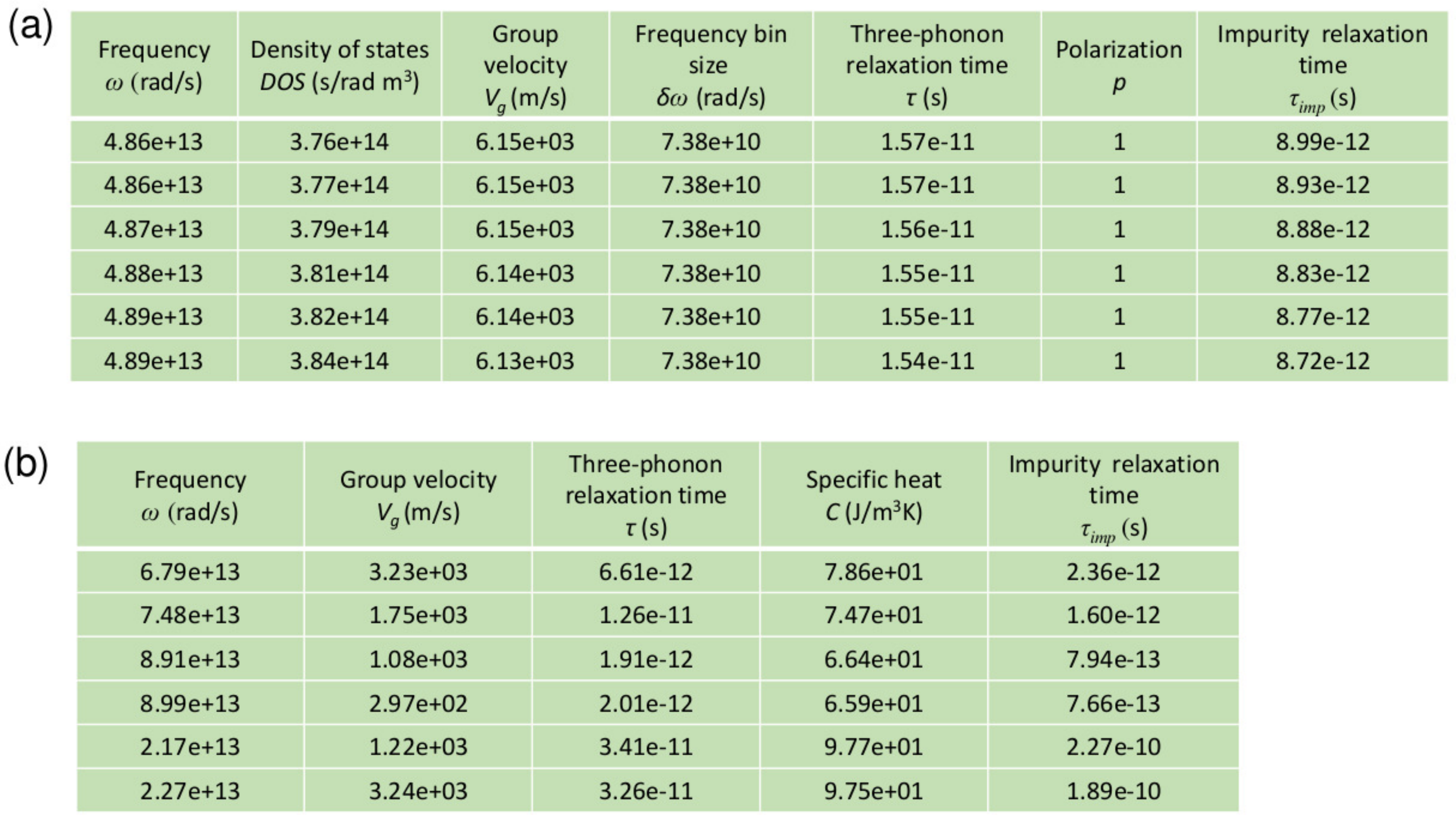}
  \cprotect\caption{(a) A snippet of the first format type of \verb|mat_data.txt| file containing $\omega$\,(rad/s), DOS\,(s/rad$\cdot$m$^3$), $V_g$\,(m/s), $\delta \omega$\,(rad/s), $\tau$\,(s), $p_i$ and, $\tau_{imp}$\,(s) in the same order. (b) Same as panel (a) but for the second format containing $\omega$ (rad/s), $V_g$ (m/s), $\tau$\,(s) , $C$\,(J/(m$^3\cdot$K)) and, $\tau_{imp}$\,(s) in the same order. See details in the text.}
  \label{fig:mat_example}
\end{figure}

Data are supplied via a file \verb|mat_data.txt|. The code supports two input formats to specify material data. The first format contains 6 columns containing $\omega_i$, density of states (DOS), $V_{g,i}$, size of frequency bin $\delta\omega_i$, $\tau_i$, and polarization $p_i$ (1 for LA and 2 for TA phonon) for the $i^{th}$ bin. $p_i$ is not required for our implementation. This format is preserved for benchmarking our development with 2-D solutions of Peraud \emph{et al.}~\cite{peraud2011efficient, peraud2012alternative, phonon2D}. The second format contains four columns containing $\omega_i$, $V_{g,i}$, $\tau$, and $C$, typically generated from post-processing of first-principles density functional theory (DFT) simulations~\cite{carrete2017almabte}. In our implementation, we treat three-phonon processes separately from two-phonon processes (impurity scattering) without using the Matthiessen rule to compute effective relaxation time. An optional column can be added in the \verb|mat_data.txt| in the end to specify impurity scattering relaxation times $\tau_{imp}$ in seconds. If this column is specified, the code uses $\tau_{imp}$ for two-phonon processes; otherwise, two-phonon processes are not considered. Figures~\ref{fig:mat_example}(a) and~(b) show a snippet of first and second format from the \verb|mat_data.txt| file.

\subsection{Geometry}

\begin{figure}
  \centering
  \includegraphics[width=\textwidth]{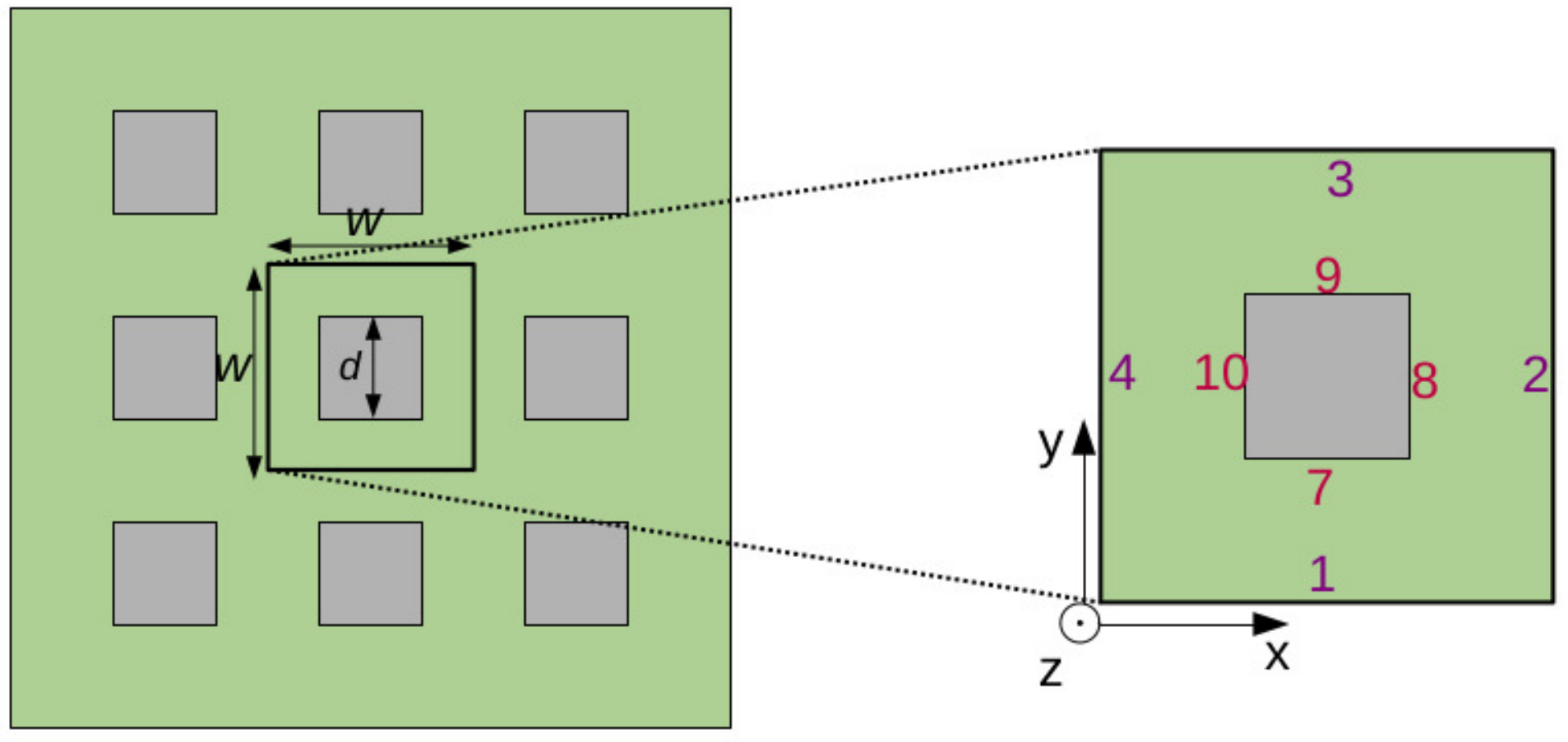}
  \caption{Periodic nanomesh geometry and unit cell. Nanomesh unit cell is shown with solid black lines with an enlarged view of the unit cell alongside. Unit cell dimension and pore size are $W = 34$\,nm and $d = 11$\,nm, respectively. The out-of-plane thickness is 22\,nm. Boundaries are marked with numbers denoting their ID. See details in the text.} 
  \label{fig:geo_example}
\end{figure}

\begin{figure}
  \centering
  \includegraphics[width=\textwidth]{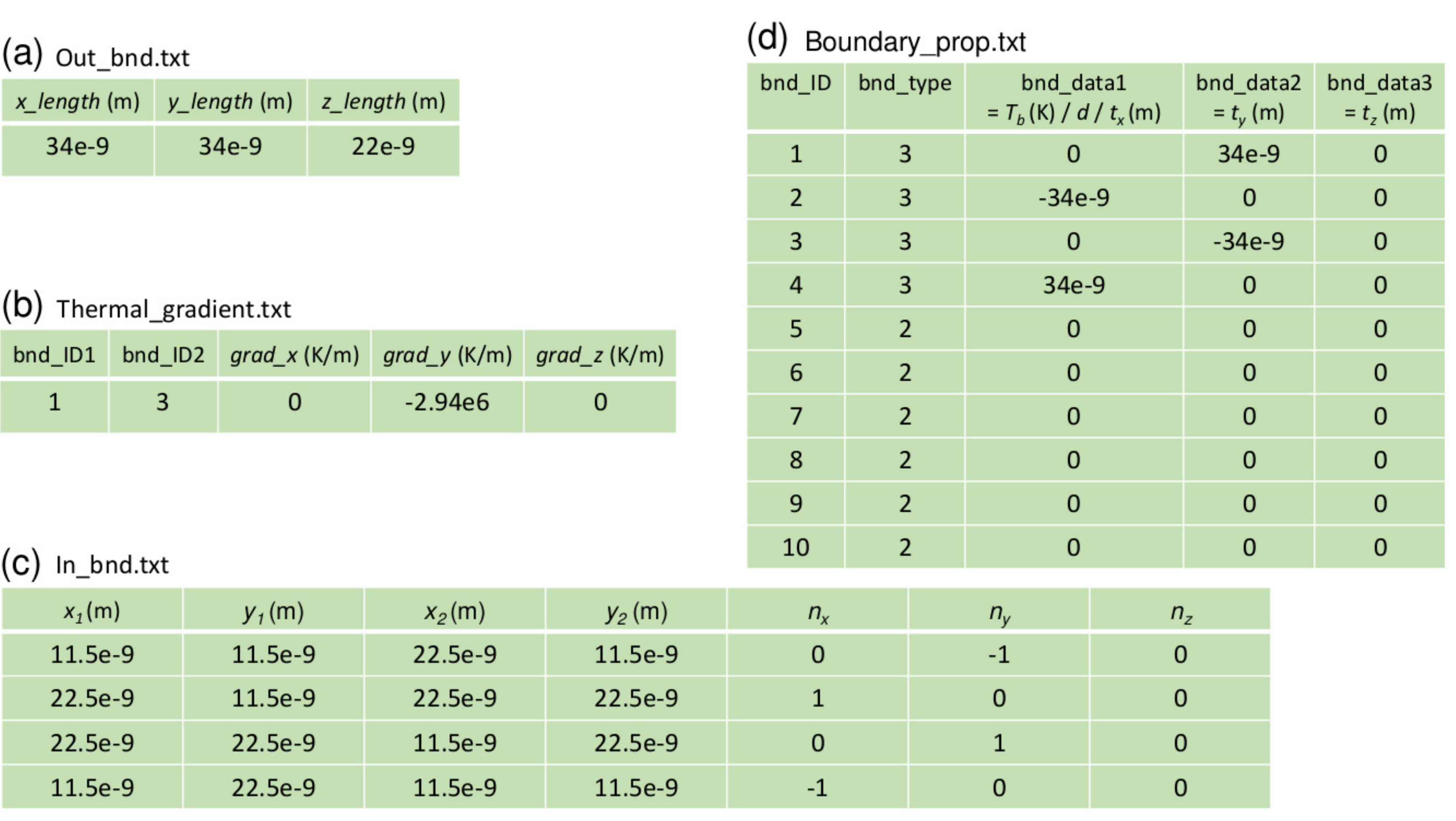}
 \cprotect\caption{Input files for specifying the geometry, boundary conditions, and source terms for the periodic nanomesh problem shown in Fig.~\ref{fig:geo_example}. (a) \verb|Out_bnd.txt| file defines the extents of the domain along $x$, $y$, and $z$-axes in units of a meter. (b) \verb|Thermal_gradient.txt| file defines the constant thermal gradient in units of K/m along the negative $y$-axis. (c) \verb|In_bnd.txt| file defines the four internal boundaries 7 to 10. (d) \verb|Boundary_prop.txt| file defines the boundary conditions (periodic/isothermal/adiabatic) for boundaries 1 to 10. See details in the text.}
  \label{fig:bnd_data}
\end{figure}

The geometry is defined using two files -- \verb|Out_bnd.txt| and \verb|In_bnd.txt|. The file \verb|Out_bnd.txt| defines the extent of a unit-cell domain in all three dimensions and contains \verb|x_length|, \verb|y_length| and \verb|z_length| of the domain in the same order in one line. We illustrate the format of the files by an example nanomesh problem (see Fig.~\ref{fig:geo_example}), also used in Sections~\ref{sub_nanomesh} and~\ref{cum_sub_nanomesh} for benchmarking our code. Figure~\ref{fig:bnd_data}(a) defines the extent of the nanomesh shown in Fig.~\ref{fig:geo_example} as required by the file \verb|Out_bnd.txt|. For simplicity, one corner of the cuboid defining the domain is assumed to lie always at the origin. \verb|In_bnd.txt| file is defined for the internal boundaries (i.e., boundaries 7 to 10) in the domain. The internal boundaries are specified by listing the end coordinates ($x_1$, $y_1$, $x_2$, $y_2$) of the line followed by the normal ($n_x$, $n_y$, $n_z$) pointing into the domain. The current version of the code assumes that the internal boundaries are perpendicular to the $x-y$ plane and extend through the thickness. Figure~\ref{fig:bnd_data}(b) shows entries of the \verb|In_bnd.txt| file.

\subsection{Boundary conditions and source term}
\verb|Boundary_prop.txt| file defines the boundary types. The format of the data is in the following order: \verb|bnd_ID|, \verb|bnd_type|, \verb|bnd_data1|, \verb|bnd_data2|, \verb|bnd_data3|. \verb|bnd_ID| 1 to 6 must be used for the outer boundaries in following order: y=0, x=\verb|x_length|, y=\verb|y_length|, x=0, z=0, z=\verb|z_length|. We consider 3-D geometry with boundaries parallel to the $x$, $y$, and $z$ axes. Internal boundaries are listed starting from \verb|bnd_ID| 7 in the same order as they appear in the \verb|In_bnd.txt| file. We consider three boundary types: 1) isothermal, 2) adiabatic and, 3) periodic. Isothermal boundary is specified using \verb|bnd_type|=1 and \verb|bnd_data1|=$T_b$.  We do not need to specify \verb|bnd_data2| and \verb|bnd_data3|, and both are entered as $0$. Adiabatic boundary is specified using  \verb|bnd_type|=2 and \verb|bnd_data1|=$d$ (degree of specularity). \verb|bnd_data2| and \verb|bnd_data3| are $0$ for adiabatic boundary. Periodic boundary is specified using \verb|bnd_type|=3. \verb|bnd_data1-3| express the $x$, $y$ and, $z$ components of the periodic translational vector $\mathbf{t}$. For example, a particle incident on \verb|bnd_ID| 1 is translated by ${\bf t} = (0, 34\times10^{-9}, 0)$\,m and is re-inserted into the domain from \verb|bnd_ID| 3. Figure~\ref{fig:bnd_data}(d) shows contents of the \verb|Boundary_prop.txt| file for boundaries 1 to 10 (see Fig.~\ref{fig:geo_example}). We define the outer boundaries of unit cell (ID 1-4) as periodic boundaries, while the inner boundaries of the pore (ID 7-10) are specified as diffusively reflecting ($d=0$). Boundaries at z=0 (ID 5) and z=\verb|z_length| (ID 6) are also taken to be diffusively reflecting. Boundary type 1 (isothermal boundary) also serves as source for the deviational particles. For periodic nanostructures, the thermal gradient is specified in the \verb|Thermal_gradient.txt| file. The format is: \verb|bnd_ID1, bnd_ID2, grad_x, grad_y, grad_z |, where \verb|bndID1| and \verb|bndID2| are the IDs of periodic boundary pair, and \verb|grad_x|, \verb|grad_y| and, \verb|grad_z| are the $x$, $y$ and, $z$ components of the thermal gradient. For example, Fig.~\ref{fig:bnd_data}(c) shows the \verb|Thermal_gradient.txt| file specifying a temperature difference of $0.1$\,K between the boundaries with ID 1 and 3, i.e., \verb|grad_y| = $-0.1/(34\times10^{-9}) = -2.94\times10^{-6}$\,K/m.

\subsection{Simulation parameters} 
\begin{figure}
  \centering
  \includegraphics[width=\textwidth]{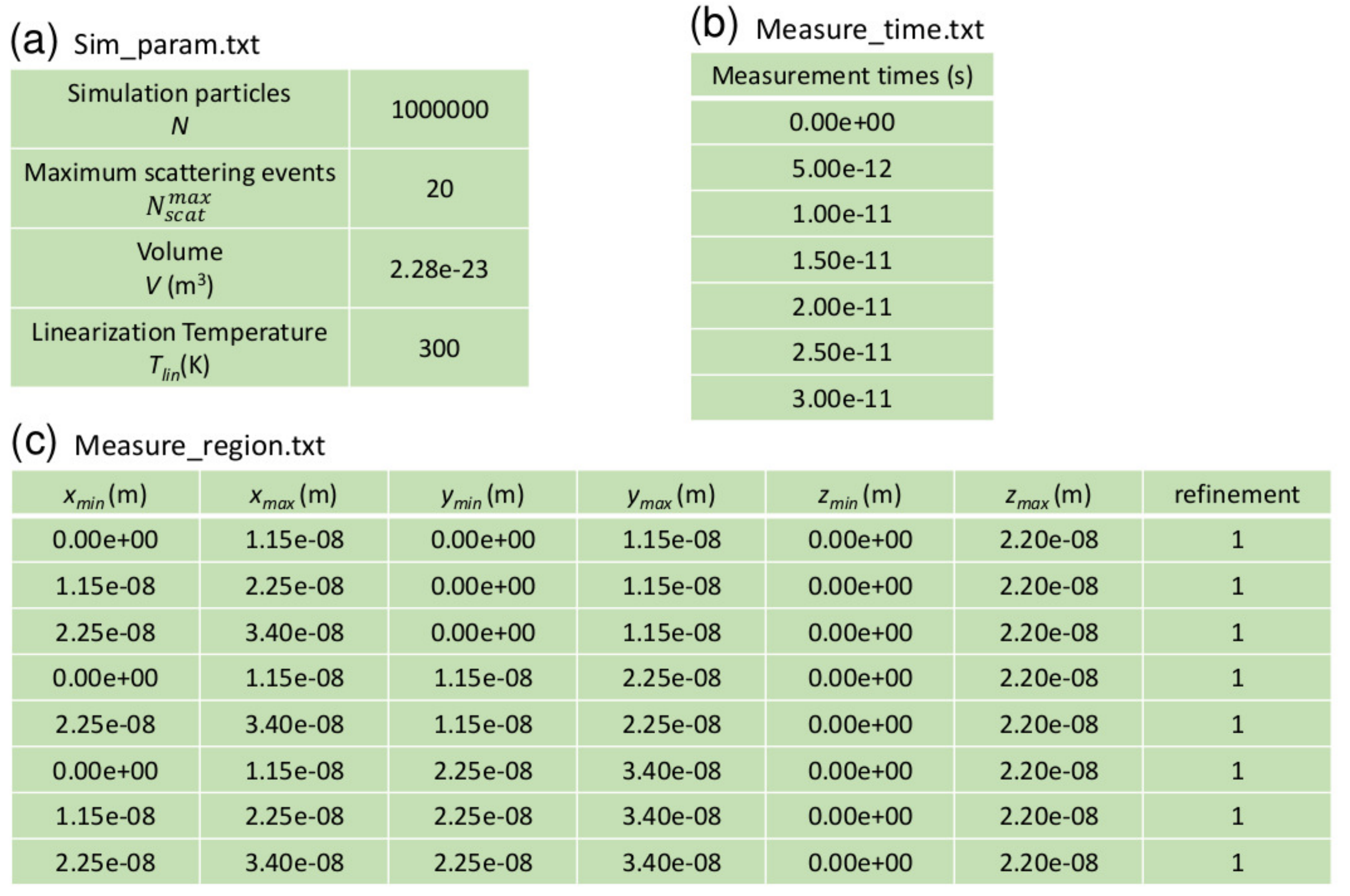}
 \cprotect\caption{Input files for specifying the simulation parameters and output requests. (a) Contents of the \verb|Sim_param.txt| for a typical simulation. (b,c) \verb|Measure_times.txt| and \verb|Measure_region.txt| files specifying the times in seconds (b) and measurement locations in meters (c) for which output is requested. See details in the text.}
  \label{fig:sim_measure}
\end{figure}

Simulation parameters are defined in \verb|Sim_param.txt| file. The file contains number of computational particles $N$, maximum number of scattering events allowed for a particle $N_{scat}^{max}$, the volume (in m$^3$) of the simulation domain $V$, and the linearization temperature $T_{lin}$ in Kelvin in the same order. For steady-state simulations, in the absence of an isothermal boundary, $N_{scat}^{max}$ must be specified to terminate the particle's trajectory. If not specified, the particle will stay in the domain indefinitely. Only three-phonon processes are counted towards $N_{scat}^{max}$ as they result in relaxation towards the equilibrium distribution. $N_{scat}^{max}$ must be large enough so that the contribution of a particle to the heat flux has converged. The volume is calculated by excluding all pores from the simulation domain. Figure~\ref{fig:sim_measure}(a) shows entries of \verb|Sim_param.txt| file for steady-state simulation of the nanomesh problem (see Fig.~\ref{fig:geo_example}): $N = 1000000$, $N_{scat}^{max} = 10$, $V = 22\times10^{-9}\times[(34\times10^{-9}\times34\times10^{-9}) - (11\times10^{-9}\times11\times10^{-9})] = 2.28\times10^{-23}$\,m$^3$, and $T_{lin} = 300$\,K.

\subsection{Output request} 
The code returns volume-averaged temperature and heat flux components for the spatial regions listed in \verb|Measure_region.txt| file at time stamps listed in \verb|Measure_time.txt| file. Figure~\ref{fig:sim_measure}(b) shows a snippet from the \verb|Measure_time.txt| file from a transient simulation. Here we request output at every $5$\,ps time interval. Figure~\ref{fig:sim_measure}(c) shows the measurement locations specified for a steady-state simulation for an example nanomesh problem (see Fig.~\ref{fig:geo_example}). We define a sampling region as a cuboid aligned with the cartesian axes. The first six entries in a row are: $x_{min}$, $x_{max}$, $y_{min}$, $y_{max}$, $z_{min}$ and, $z_{max}$ for the cuboid in units of meter. The last entry specifies further refinement of the region into smaller cuboids. For a given $n$, the region is further divided into $2^{3n}$ equal regions. We report the divided regions in the \verb|detector_location.txt| file in the same format as in the \verb|Measure_region.txt| with six entries defining the location of a `detector.' The output is reported in the \verb|T{}.txt|, \verb|Qx{}.txt|, \verb|Qy{}.txt| and \verb|Qz{}.txt| files, where \verb|{}| denotes the equilibrium temperature at which simulation is performed. Each row of the output file corresponds to the detector located in the \verb|detector_location.txt|. For transient simulations, each output column corresponds to the time stamps defined in the \verb|Measure_time.txt|. For steady-state simulations, each output column corresponds to the frequency bins defined in the \verb|mat_data.txt|. The last described output format is particularly useful in calculating the cumulative thermal conductivity of nanostructures. We note that in the \verb|T{}.txt| file, the deviation of temperature form the equilibrium baseline value rather than the true temperature is reported.

\subsection{Instructions to run the program}
The program can be executed either on a single node with multiple processors sharing the same memory using \verb|Single_node_multiple_proc.m| or on multiple nodes using \verb|Distributed_computing.m|. Both of the files are available in the \verb|example_input_files| directory at the GitHub repository. The \verb|MATLAB| package requires access to \verb|Parallel Computing Toolbox| for execution and can be run from GUI or command-line. An open-source alternative, an \verb|Octave| implementation, is also provided at the GitHub repository in \verb|Octave_implementation| directory. The \verb|Octave| program can either be run from GUI or on command-line using \verb|octave --persist BTE_solution_3D.m|.

\section{Illustrative examples} \label{sec_bench}
To demonstrate the functionality and ascertain the accuracy of our code's output, we provide several illustrative example problems for which either a theoretical/analytical solution exists or published computational results are available.

\subsection{Ballistic heat conduction} \label{sub_transient_1d}

\begin{figure}
\begin{center}
  \includegraphics[width=0.7\textwidth]{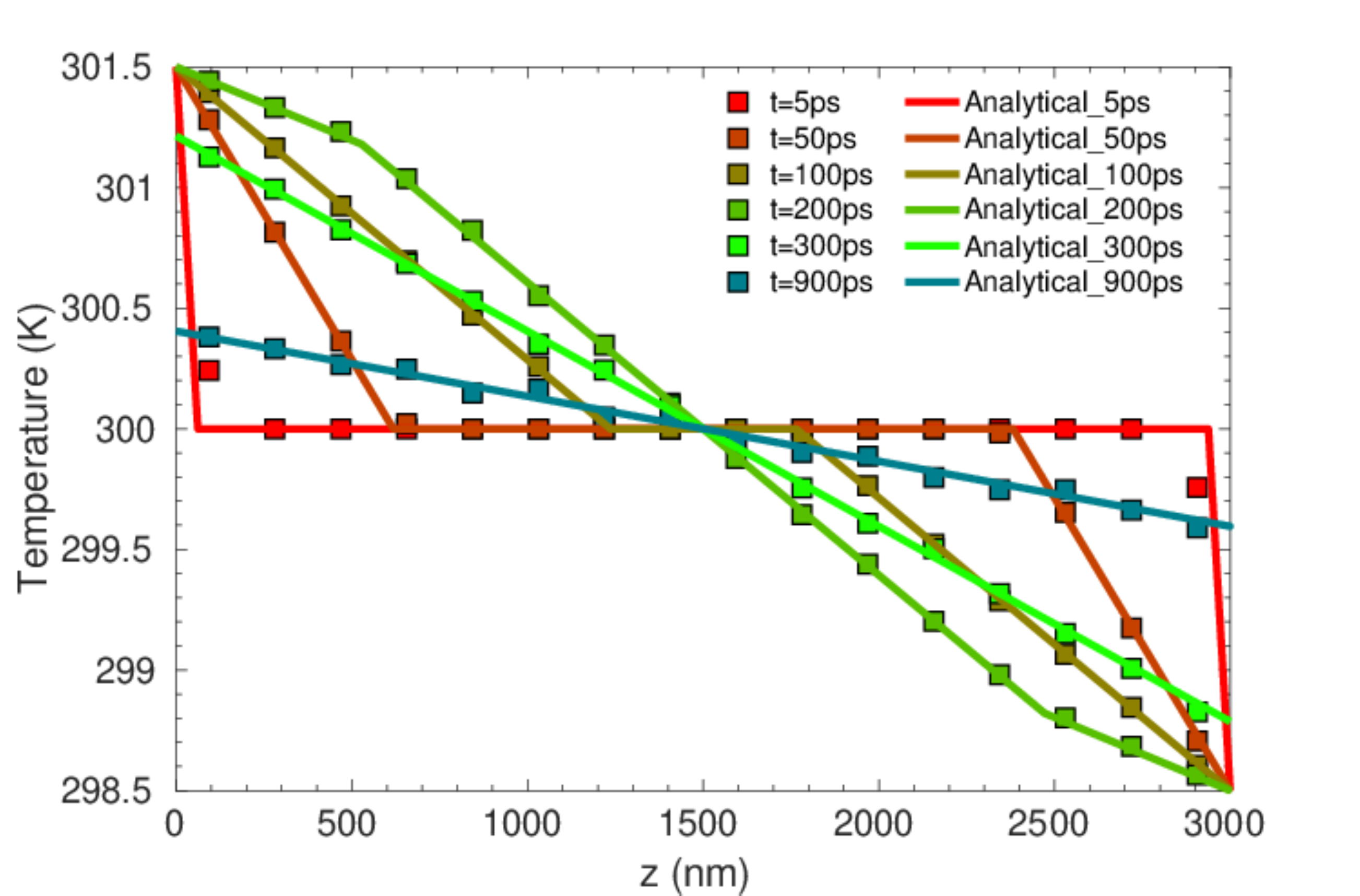}
  \end{center}
  \caption{Transient ballistic heat conduction. A comparison of analytical (Eq.~\eqref{eq:Analytical_T}, solid lines) and LBTE solution (filled markers) of temperature along the $z$-axis at $t = $ 5, 50, 100, 200, 300, and 900\,ps.}
  \label{fig:ballistic_1d}
\end{figure}

We perform a simulation of ballistic 1D heat transfer using the Debye model for phonon dispersions and fixed temperature at the two ends. Analytical solution of temperature deviation from equilibrium is given by~\cite{peraud2011efficient}
\begin{align}
  \label{eq:Analytical_T}
  \Delta T(z,t) = \frac{1}{2} \left ( 1 - \frac{z}{V_g t} \right ) &H \left( 1 - \frac{z}{V_g t}\right ) \Delta T_l \nonumber\\
  &+ \frac{1}{2} \left ( 1 - \frac{L-z}{V_g t} \right ) H \left( 1 - \frac{L-z}{V_g t}\right ) \Delta T_r,
\end{align}
where, $H$ is the Heaviside function, $L$ is the length of the domain, $V_g$ is the phonon group velocity, $\Delta T_l$ and $\Delta T_r$ are deviation from equilibrium temperature for \emph{left} and \emph{right} walls. To simulate the ballistic conduction with our code, we choose a 3D domain of $3000$\,nm$\times 3000$\,nm$\times 3000$\,nm. At $t = 0$, the wall temperature at $z$ = 0 and $z$ = 3000\,nm is impulsively set to 303 and 297\,K, respectively. We apply periodic boundary conditions at $x$ = 0 and 3000\,nm, and $y$ = 0 and 3000\,nm. $V_g$ and $\tau$ are taken to be 12360\,m/s and 1\,s, respectively. A large value of $\tau$ prevents any three-phonon processes within the simulation duration, a necessary requirement for the ballistic heat conduction. Figure~\ref{fig:ballistic_1d} compares the temperature along the $z$-axis at different times with the analytical expression, showing an excellent agreement. Here we bring attention to the discontinuities at the boundary, i.e., for $t>200$\,ps, the domain temperature near the boundary is less than the boundary temperature. In the ballistic limit, at steady-state, the temperature within the domain would assume a constant value of $[(T_L^4+T_R^4)/2]^{1/4}$ governed by the Stefan-Boltzmann law~\cite{mazumder2001monte}, where $T_L$ and $T_R$ are the left and right boundary temperatures, respectively.

\subsection{Quasi-ballistic heat conduction and comparison with the Fourier law}

\begin{figure}
\begin{center}
  \includegraphics[width=0.7\textwidth]{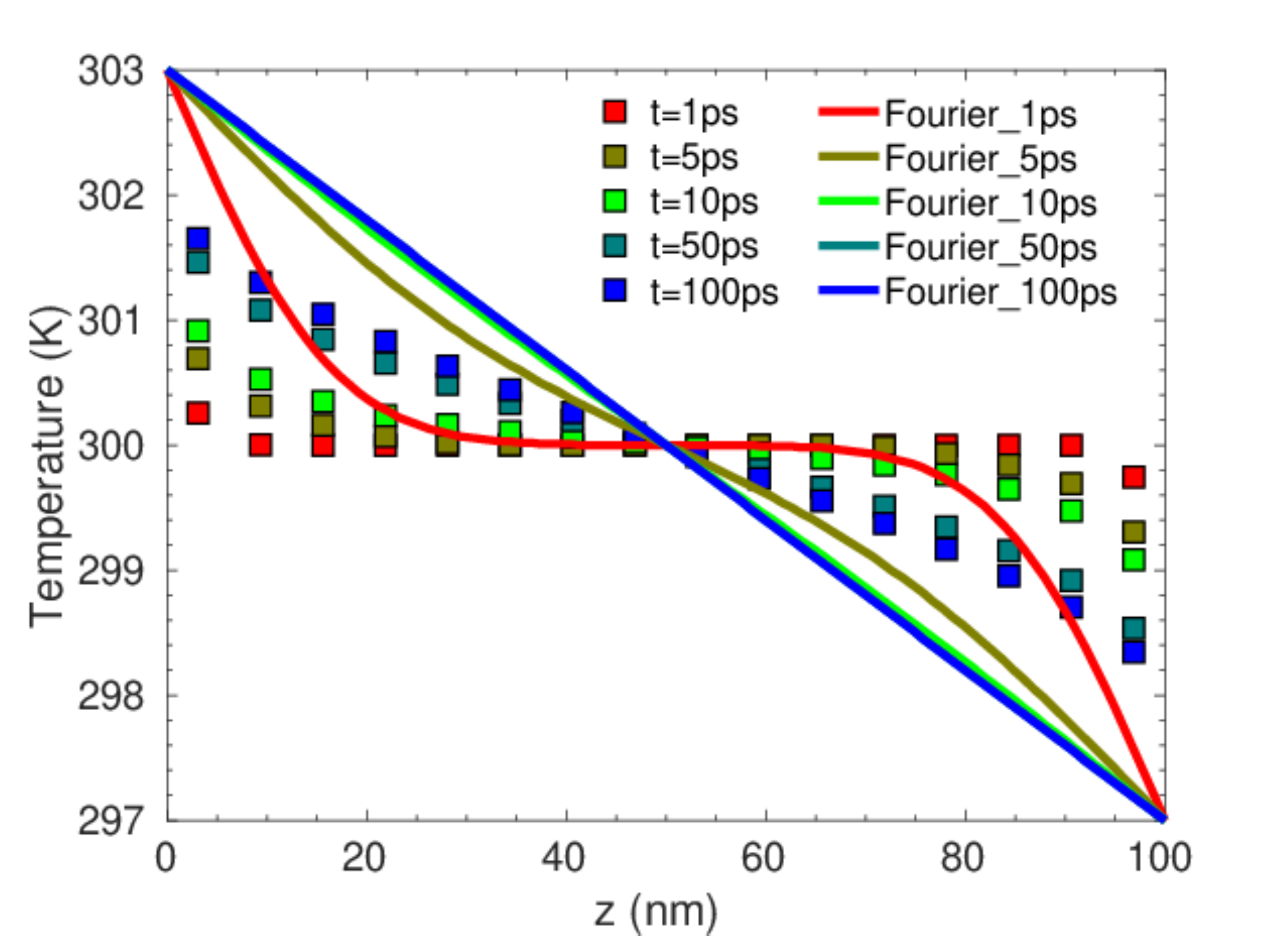}
  \end{center}
  \caption{Transient quasi-ballistic heat conduction. A comparison of temperature along the $z$-axis obtained using the LBTE (filled markers) and the Fourier law of heat conduction (solid lines) at $t = $ 1, 5, 10, 50, and 100\,ps.}
  \label{fig:Fourier_compare}
\end{figure}

An excellent agreement in the ballistic limit prompted us to compare the LBTE solution with the Fourier law of heat conduction in the quasi-ballistic limit. We consider silicon at 300\,K as an example. At $t=0$, the temperature of the left and right walls are set to $303$ and 297\,K, respectively. 1D Fourier heat conduction equation is solved for a $100$\,nm domain. The initial temperature is assumed to be constant across the domain and kept at 300\,K. The bulk thermal conductivity, density, and heat capacity of silicon at 300\,K are taken as $143.84$\,W/(m.K), $2532.59$\,kg/m$^3$, and $700$\,J/(m$^3$.K), respectively. To simulate the same problem with our code, we choose a simulation domain of $100$\,nm$\times100$\,nm$\times100$\,nm. The wall temperature at $z$ = 0 and $z$ = 100\,nm is set to 303 and 297\,K, respectively. We apply periodic boundary conditions at $x$ = 0 and 100\,nm, and $y$ = 0 and 100\,nm. Per se, this is not a benchmark problem. Since Fourier law is only applicable for diffusion-like conduction, as expected, it deviates significantly from the LBTE solution in this limit, as we show in Figure~\ref{fig:Fourier_compare}. Fourier solution reaches the equilibrium temperature profile within 10's of ps. On the other hand, because of quasi-ballistic heat conduction, the LBTE solution lags.

\subsection{Thermal conductivity of a thin-film} \label{sub_steady_2d}

\begin{figure}
\begin{center}
  \includegraphics[width=\textwidth]{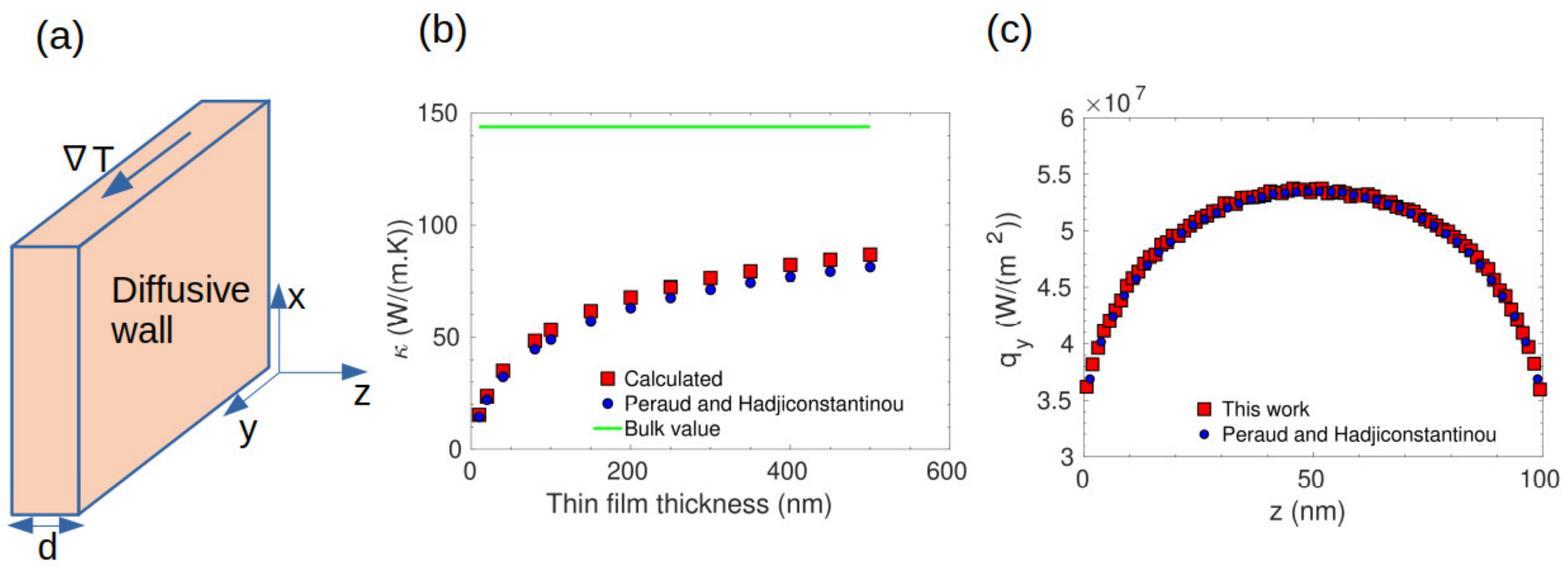}
  \end{center}
  \caption{Thermal conductivity ($\kappa$) and heat flux of a thin-film of silicon. (a) Simulation geometry -- the domain is infinite in $x$ and $y$ directions, and $d$ is varied from 20 to 500\,nm. (b,c) Comparison of $\kappa$ (b) and heat flux q$_y$ (c) from our LBTE solution with the results of Peraud and Hadjiconstantinou~\cite{peraud2011efficient}. For panel (c), $d = 100$\,nm and the equilibrium temperature is 300\,K. Simulations are steady-state.}
  \label{fig:thin_conduct}
\end{figure}

We calculate the thermal conductivity ($\kappa$) of a thin film of thickness $d$ at $T=300$\,K for an applied temperature gradient along the $y$-axis, as shown in Fig.~\ref{fig:thin_conduct}a. This problem has been solved analytically and computationally by Peraud and Hadjiconstantinou for silicon~\cite{peraud2011efficient}. To calculate $\kappa$ and heat flux with our code, we choose a simulation domain of $100$\,nm$\times100$\,nm$\times d$\,nm, where $d$ varies from 20 to 500\,nm. The boundaries at $z = 0$ and $z = d$ are modeled as diffusive walls, while periodic boundary condition is applied $x$ = 0 and 100\,nm, and $y$ = 0 and 100\,nm. A thermal gradient of $-5 \times 10^5$\,K/m is applied along the $y$-axis. We calculate the steady-state temperature and heat-flux in the simulation domain. Figure~\ref{fig:thin_conduct}(b,c) compares $\kappa$ and $y$ component of the heat flux ($q_y$) from our LBTE solution with the results of Ref.~\citenum{peraud2011efficient} showing an excellent agreement. We note that for few values of $d$, such as $d=450$\,nm, we require a large number of computational particles (8 million as opposed to 1 million) to obtain acceptable noise levels in the $\kappa$. This is due to a large contribution of phonons to the heat flux that has a low density of states, as also noted by Peraud~\cite{peraud2015efficient}. Due to the low density of states, these phonons are sampled less frequently than the others, and their contribution to heat flux (and consequently $\kappa$) has a large variance, leading to fluctuations in calculated $\kappa$.  

\subsection{Thermal conductivity of a thin-film using first-principles DFT data} \label{sub_DFT}

\begin{figure}
  \centering
  \includegraphics[trim=4.15cm 0.75cm 4.45cm 0.80cm, clip=true, width=1.00\textwidth]{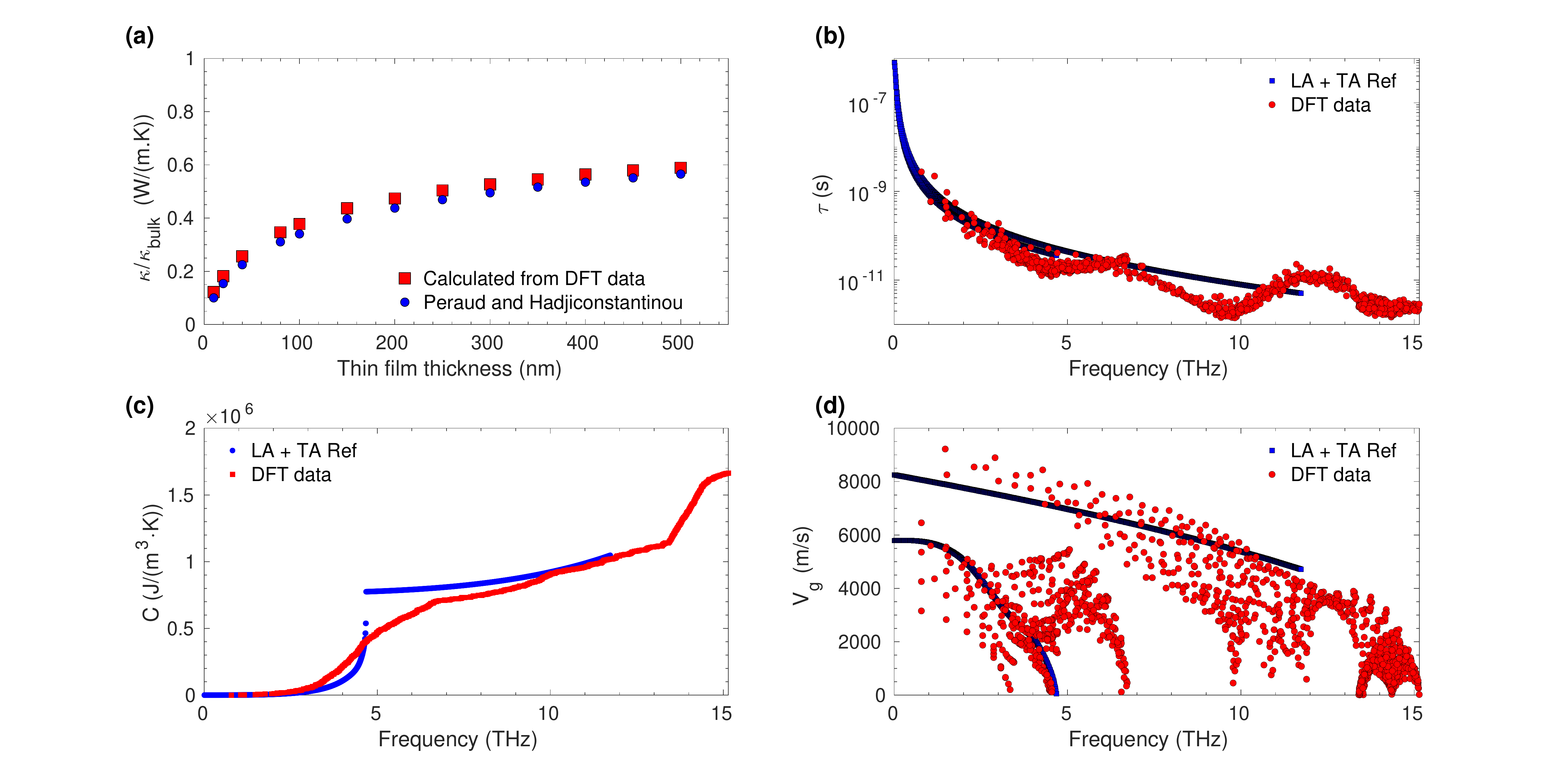}
  \cprotect\caption{Thermal conductivity ($\kappa$) of a thin-film of silicon calculated using first-principles DFT data. (a) $\kappa/\kappa_{bulk}$ from our LBTE solution as a function of  thin film thickness compared with simulations of Peraud and Hadjiconstantinou~\cite{peraud2011efficient}. (b,c,d) A comparison of first-principles DFT data of $\tau$ (b), heat capacity (c), and $V_g$ (d) with the reference \verb|LA + TA Ref| data. See details in the text.}
  \label{fig:DFT}
\end{figure}

In the previous Section~\ref{sub_steady_2d}, phonon dispersions of longitudinal and transverse acoustic (LA and TA) branches were assumed to be isotropic and optic branches were not included. $\omega$, $V_g$, and $\tau$ were calculated using the empirical models such that the calculated $\kappa$ is in agreement with the bulk $\kappa$ of silicon. As mentioned earlier, this is the same data as used by Peraud and Hadjiconstantinou~\cite{peraud2011efficient}, and hereafter we refer it to as \verb|LA + TA Ref|. However, it is well known that the phonon dispersion of silicon is highly anisotropic~\cite{kim2018nuclear}. Here we calculate $\omega$, $V_g$, and $\tau$ from first-principles density functional theory (DFT) simulations of silicon. We use the same raw data as reported in Ref.~\citenum{carrete2017almabte} from the almaBTE database. A comparison of frequency-resolved $\tau$, heat capacity $C$, and $V_g$ from first-principles DFT simulations with \verb|LA + TA Ref| data is shown in Fig.~\ref{fig:DFT}(b-d). As one can observe, although the order of magnitude is generally agreeable, differences are evident. We use frequency-resolved first-principles DFT data to calculate $\kappa$ as a function of a thin-film thickness of $d$. Figure~\ref{fig:DFT}(a) compares the thin film $\kappa$ normalized to $\kappa_{bulk}$ from our LBTE solution using first-principles DFT data with the results of Ref.~\citenum{peraud2011efficient}. The results are in good agreement. A small overestimation is attributed to the variation in frequency-resolved input datasets. 

\subsection{Thermal conductivity of nanomesh} \label{sub_nanomesh}

\begin{figure}
  \begin{center}
    \includegraphics[width=0.65\textwidth]{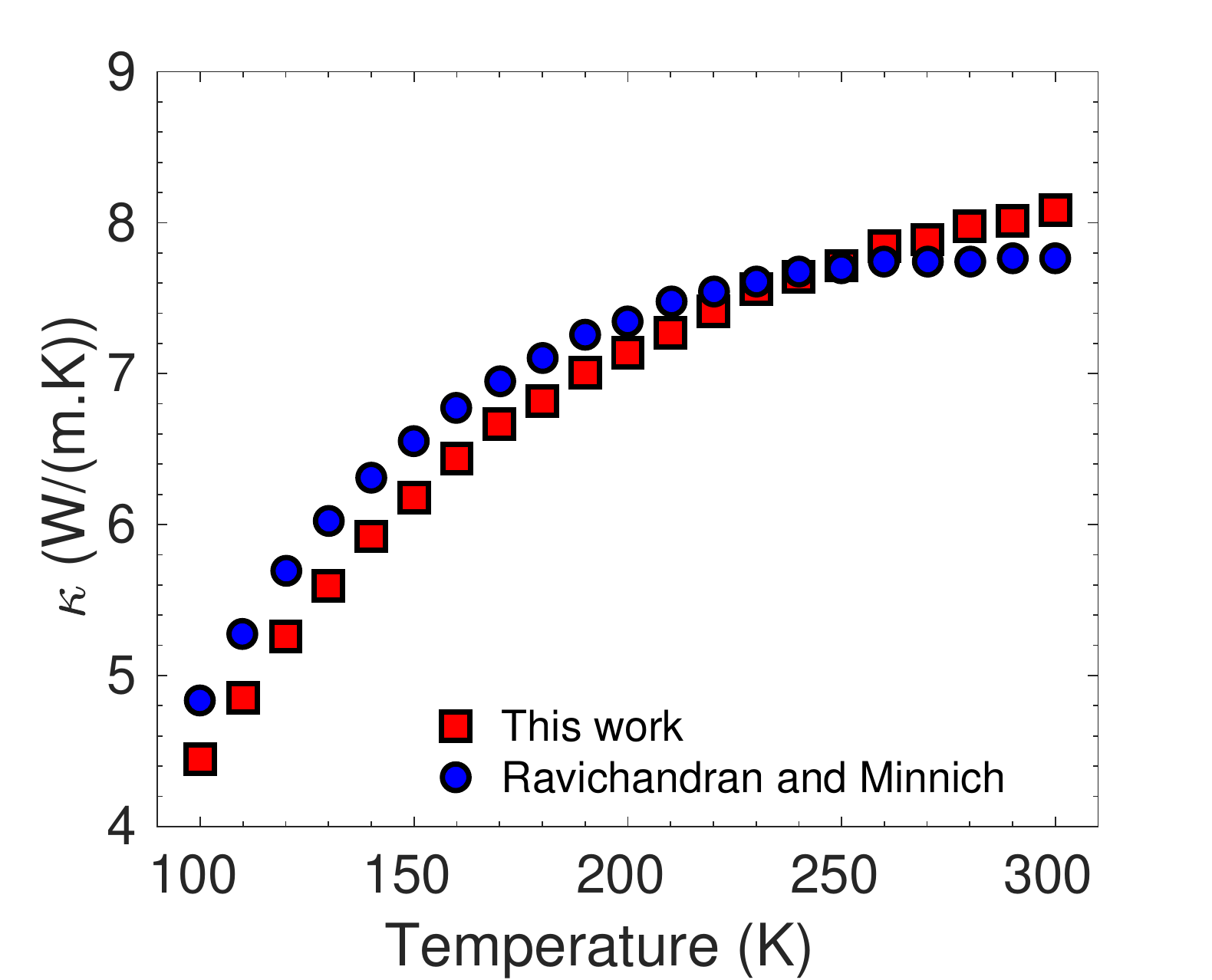}
  \end{center}
  \caption{A comparison of $\kappa$ from our LBTE solution with simulations of Ravichandran and Minnich~\cite{ravichandran2014coherent} for the geometry shown in Figure~\ref{fig:geo_example}.}
  \label{fig:nanomesh_compare}
\end{figure}

We calculate the thermal conductivity of a periodic nanomesh of silicon. This problem has been solved numerically by Ravichandran and Minnich~\cite{ravichandran2014coherent}. The unit cell is shown in Fig.~\ref{fig:geo_example}. Unit cell is square in the $x-y$ plane with $W = 34$\,nm. In the out-of-plane direction, the thickness is 22\,nm. The pore is at the center of the unit cell ($d = 11$\,nm) and extends throughout the thickness. To model the silicon nanomesh with our code, we apply a thermal gradient corresponding to a temperature difference of 0.1\,K along the $y$-axis of the unit cell. We define the outer boundaries of the unit cell (ID 1-4) as periodic boundaries, while the inner boundaries of the pore (ID 7-10) are specified as diffusively reflecting ($d=0$). Boundaries at z=0 (ID 5) and z=\verb|z_length| (ID 6) are also taken to be diffusively reflecting. More details are described in Section~\ref{sec_code}. An average heat flux across the $y$ direction is calculated and is divided by the thermal gradient to calculate the $\kappa$. To facilitate a direct comparison with the results of Ref.~\citenum{ravichandran2014coherent}, we include the same frequency-dependent impurity scattering given by $ \tau_{imp}^{-1} = 2 \times 10^{-44} \omega^4$ s$^{-1}$. Figure~\ref{fig:nanomesh_compare} compares the LBTE solution with the simulation of Ref.~\citenum{ravichandran2014coherent} from $100\leqslant T\leqslant 300$\,K. The results are in reasonable quantitative agreement. A small discrepancy throughout the temperature is expected since silicon phonon dispersions and three-phonon relaxation time from Ref.~\citenum{ravichandran2014coherent} are not available to us. Instead, here we use the data from Peraud and Hadjiconstantinou~\cite{peraud2011efficient}.

\subsection{Frequency-resolved cumulative thermal conductivity of nanomesh} \label{cum_sub_nanomesh}

\begin{figure}
  \begin{center}
    \includegraphics[width=0.65\textwidth]{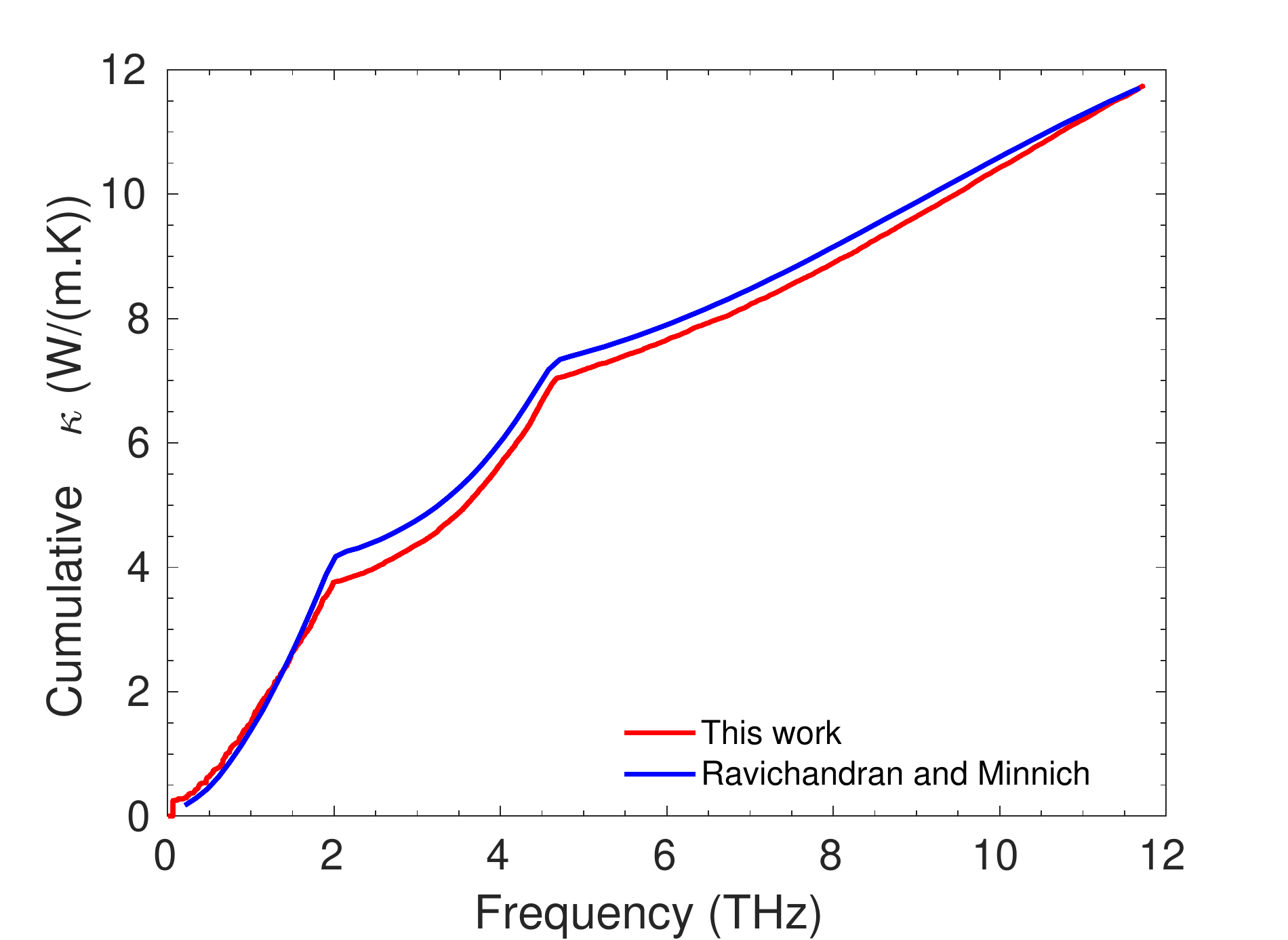}
  \end{center}
  \caption{A comparison of frequency-resolved cumulative $\kappa$ of silicon nanomesh at 300 K from our LBTE solution with simulations of Ravichandran and Minnich~\cite{ravichandran2014coherent} ($\leqslant2$\,THz phonons are specularly reflected). See details in the text.}
  \label{fig:nanomesh_accumu}
\end{figure}

In literature, $\kappa$ is often spectrally resolved to calculate the relative contribution of different phonon frequencies. Ravichandran and Minnich~\cite{ravichandran2014coherent} had calculated the frequency-resolved cumulative $\kappa$ for silicon nanomesh. The only difference from $\kappa$ simulations of Section~\ref{sub_nanomesh} is that the phonons of energy less than 2\,THz are specularly reflected, which increases the $\kappa$ from $\sim$8 to $\sim$12\,W/m/K at 300\,K. Figure~\ref{fig:nanomesh_accumu} compares the frequency-resolved cumulative $\kappa$ of our simulation with Ref.~\citenum{ravichandran2014coherent} at 300 K. The results are in quantitative agreement. A small difference is attributed to the different silicon data used in our simulations, as described earlier in Section~\ref{sub_nanomesh}.

\section{Parallelization} \label{sub_parallel}

\begin{figure}
 \begin{center}
  \includegraphics[width=\textwidth]{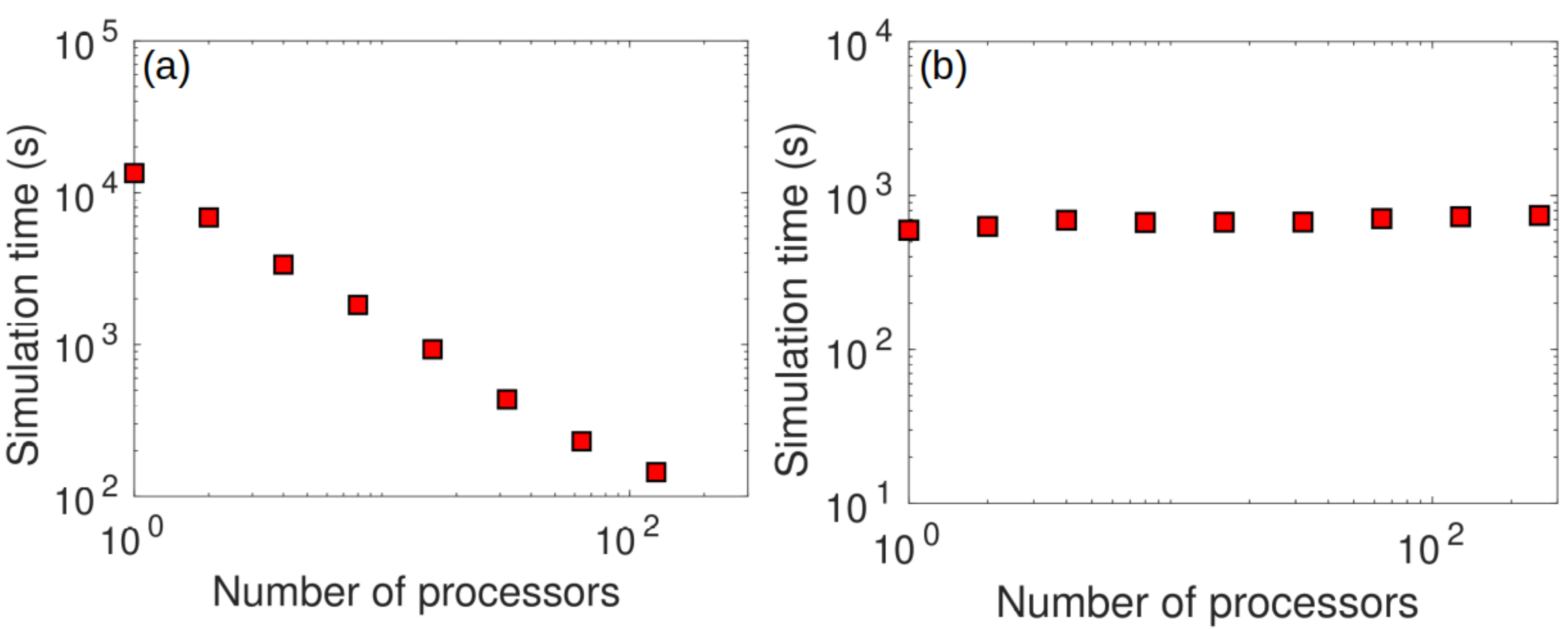}
   \end{center}
  \caption{Scaling performance of MC solution of LBTE. (a) Scaling with constant problem size having $N = 1,000,000$ computational particles. (b) Scaling with constant load of $n = 50,000$ computational particles on each processor.}
  \label{fig:parallel_perform}
\end{figure}

The advantage of LBTE is that particle trajectories are independent of each other, making implementation embarrassingly parallel. To run particle trajectories in parallel, we write our code using the Single Program Multiple Data (SPMD) methodology of parallelization. We test our code on the Matlab Distributed Computing Server (MDCS). Although the code does not impose any inherent limit on the number of processors, we restrict our simulations to 256 processors. We perform two tests to access scalability performance on a problem described in Section~\ref{sub_steady_2d} for $d = 100$\,nm. In the first test, we keep the total number of computational particles fixed at $N=1,000,000$. We increase the number of processors from one processor to 256 processors. Simulation time as a function of the number of processors is shown in Figure~\ref{fig:parallel_perform}a. A near-linear trend highlights that simulation time can be decreased by increasing the number of processors provided I/O overhead is not excessive. The second test is performed by increasing the problem size but keeping the load constant on each processor at $n = 50,000$ computational particles. It is reassuring to see in Fig.~\ref{fig:parallel_perform}b that $N$ (= $n\times$number of processors) could be increased to improve the estimate of thermodynamic observable without much increase in simulation time if more processors become available.

\section{Summary} \label{sec_summary}
The open-source MCBTE program presented here simulates the LBTE using the Monte-Carlo solution approach. In our implementation, we can specify equilibrium temperature to be a constant value in the entire domain or can have a constant gradient along one or more directions. The latter is very useful in calculating $\kappa$ that can further be integrated with the multiphysics simulation~\cite{nardi2011probing, hoogeboom2015new}. The near-linear scaling on parallelization provides the opportunity to simulate large domains for longer time durations. Moreover, the source code can be modified with minimal changes/additions to simulate the problem of interest. For example, various geometries to simulate the effect of size, patterns, and periodicity of nanostructures and nanocomposties on the thermal transport can be easily studied as illustrated in Section~\ref{sub_nanomesh}. The frequency-resolved output of the heat flux provides the relative contribution of phonons. This spectral information can be used to enhance/suppress the thermal transport by effectively tuning phonons' reflection properties from coherent to incoherent or vice-versa, as demonstrated in Section~\ref{cum_sub_nanomesh}. Interface scattering can be incorporated by considering frequency- and/or angle-of-incidence-dependent transmission probabilities, which can further be extended to study thermal transport in polycrystalline material by defining grain-boundaries as interfaces and resampling the scattered phonon. We are currently developing a user-friendly interface for the interface scattering. The calculated interface properties can then be an input to the finite element analysis of the continuum model to explain experimental observations, such as ultrafast pump-probe measurements of quasi-ballistic thermal transport from nanoscale interfaces in fused silica and sapphire substrate~\cite{siemens2010quasi}. Moreover, the simulation of other experimental setups such as time-domain thermoreflectance~\cite{jiang2018tutorial} and spatially periodic free-standing membranes~\cite{johnson2013direct} require minor changes to include phonon generation from external heat sources.

\section*{Acknowledgements}
A. Pawnday and A.P.R. acknowledges the financial support from IRCC-IITB. D.B. thanks the financial support from DST under the project no.: SRG/2019/001238, and MHRD-STARS under the project no.: STARS/APR2019 /PS/345/FS. Authors acknowledge the use of computing resources provided by the Center for Computational Research.

\pagebreak
\bibliographystyle{elsarticle-num}

\end{document}